\newcommand*{\ATLASLATEXPATH}{latex/}

\documentclass[11pt,fleqn,UKenglish,texlive=2011,txfonts,cernpreprint]{\ATLASLATEXPATH atlasdoc} 

\pdfoutput=1

\usepackage[subfigure=true,biblatex=false]{\ATLASLATEXPATH atlaspackage}
\usepackage{\ATLASLATEXPATH atlasphysics}
\usepackage{\ATLASLATEXPATH atlasbsm}

\usepackage{subfigure}
\usepackage[T1]{fontenc}
\usepackage{slashed}
\usepackage{multirow}
\usepackage{rotating}
\usepackage{hhline}
\usepackage{lscape}
\usepackage{xspace}
\usepackage{hepnicenames} 
\usepackage{units}
\usepackage[hyphens]{url}
\usepackage[normalem]{ulem} 
\usepackage[maxfloats=30]{morefloats}
\usepackage{cite}
\usepackage{caption} 
\usepackage{siunitx}
\usepackage{color}
\hypersetup{anchorcolor=blue}

\renewcommand{\PSe} {\ensuremath{\tilde e \xspace} }
\renewcommand{\PSgm} {\ensuremath{\tilde \mu \xspace} }

\newcommand{\PSgti} {\ensuremath{\tilde \tau_1 \xspace} }

\renewcommand{\ttbar}{\antibar{t}}

\renewcommand{\met}{\ensuremath{E_{\mathrm{T}}^{\mathrm{miss}}}\xspace}

\newcommand{\ptmissv}{\ensuremath{\vec{{p}}_\mathrm{T}^\mathrm{~miss}}\xspace}
\newcommand{\meff}{\ensuremath{m_{\mathrm{eff}}}\xspace}

\renewcommand{\pT}{\pt\xspace}
\newcommand{\tauPt}{\ensuremath{\pT^{\tau}}\xspace}

\newcommand{\njet}{\ensuremath{N_\text{jet}}\xspace}
\newcommand{\nbjet}{\ensuremath{N_{b\text{-jet}}}\xspace}

\newcommand{\mass}[1]{\ensuremath{m_{#1}}}
\newcommand{\mT}{\ensuremath{\mass{\mathrm{T}}}\xspace}
\newcommand{\mTtau}{\ensuremath{\mass{\mathrm T}^{\tau}}\xspace}

\newcommand{\mTl}{\ensuremath{m^\ell_{\mathrm{T}}}\xspace}

\newcommand{\mTTT}{\ensuremath{\mass{\mathrm T}^{\tau_1}+\mass{\mathrm T}^{\tau_2}}\xspace}

\newcommand{\mohkerned}{\ensuremath{m_{{1}\kern-0.08em/\kern-0.07em{2}}}\xspace}

\renewcommand{\HT}{\ensuremath{H_{\mathrm{T}}}\xspace}

\newcommand{\onetau}{\mbox{$1\tau$}\xspace}
\newcommand{\twotau}{\mbox{$2\tau$}\xspace}

\renewcommand{\GeV}[1]{\ensuremath{#1 \gev} }
\renewcommand{\TeV}[1]{\ensuremath{#1 \tev} }

\newcommand{\inmath}[1]{\ensuremath{#1}\xspace}

\newcommand{\mttwo}{\inmath{m_\text{T2}}}

\newcommand{\mTSum}{\inmath{\mT^\mathrm{sum}}}

\skipbeforetitle{100pt}

\AtlasTitle{
  Search for squarks and gluinos in events with hadronically decaying tau leptons, jets and missing transverse momentum
 in proton--proton collisions at $\sqrt{s}=13$~\tev\ recorded with the ATLAS detector
  }  

\usepackage{authblk}

\renewcommand\Authands{, } 
\renewcommand\Affilfont{\itshape\small} 

\author{The ATLAS Collaboration}

\AtlasRefCode{SUSY-2016-01}
\PreprintIdNumber{CERN-EP-2016-145}

\AtlasJournal{EPJC}

\definecolor{lgray}{rgb}{0.75, 0.75, 0.75}

\abstracttext{
A search for supersymmetry in events with large missing transverse momentum, jets, and at least one hadronically decaying tau lepton has been performed using 3.2~fb$^{-1}$ of proton--proton collision data 
at $\sqrt{s}=13\tev$ recorded by the ATLAS detector at the Large Hadron Collider in 2015.
Two exclusive final states are considered, with either exactly one or at least two tau leptons.
No excess over the Standard Model prediction is observed in the data.
Results are interpreted in the context of gauge-mediated supersymmetry breaking and a simplified model of gluino pair production with tau-rich cascade decays, substantially improving on previous limits.
In the GMSB model considered, supersymmetry-breaking scale ($\Lambda$) values below $92 \tev$ are excluded at the 95\% confidence level, corresponding to gluino masses below $2000 \gev$.
For large values of $\tan\beta$, values of $\Lambda$ up to $107 \tev$ and gluino masses up to $2300 \gev$ are excluded.
In the simplified model, gluino masses are excluded up to $1570 \gev$ for neutralino masses around $100 \gev$.
Neutralino masses up to $700 \gev$ are excluded for all gluino masses between $800 \gev$ and $1500 \gev$, while the strongest exclusion of $750 \gev$ is achieved for gluino masses around $1400 \gev$.
}

\begin{document}

\renewcommand{\textfraction}{0.07}
\widowpenalty=10000
\clubpenalty=10000

\maketitle

\tableofcontents

\newpage

\section{Introduction}
\label{sec:intro}
Supersymmetry (SUSY)~\cite{Golfand:1971iw,Volkov:1973ix,Wess:1974tw,Wess:1974jb,Ferrara:1974pu,Salam:1974ig}
introduces a symmetry between fermions and bosons, resulting in a SUSY
partner (sparticle) for each Standard Model (SM) particle, with
identical mass and quantum numbers, and a difference of half a unit
of spin. 
Squarks ($\tilde q$), gluinos ($\tilde g$), sleptons ($\tilde \ell$) and sneutrinos ($\tilde \nu$) are the superpartners of the quarks, gluons, charged leptons and neutrinos, respectively.
The SUSY partners of the gauge and Higgs bosons are called gauginos and higgsinos, respectively.
The charged electroweak gaugino and higgsino states mix to form charginos
($\tilde\chi_{i}^{\pm}$, $i$ = 1,2), and the neutral states mix to form neutralinos
($\tilde\chi_{j}^{0}$, $j$ = 1,2,3,4).
Finally, the gravitino ($\tilde G$) is the SUSY partner of the graviton.
As no supersymmetric particle has been observed, 
SUSY must be a broken symmetry. In general, the minimal supersymmetric Standard Model (MSSM) allows couplings which violate baryon- and lepton-number conservation, leading to, for example, a short proton lifetime. 
To ensure accordance with established measurements, $R$-parity~\cite{Fayet:1976et,Fayet:1977yc,Farrar:1978xj,Fayet:1979sa,Dimopoulos:1981zb} conservation is often assumed.
In this scenario, sparticles are produced in pairs and decay through cascades involving SM particles and other sparticles until the lightest sparticle (LSP), which is stable, is produced.

Final states with tau leptons are of particular interest in SUSY searches, although they are experimentally challenging.
Light sleptons could play a role in the co-annihilation of 
neutralinos in the early universe, and models with light scalar taus are consistent
with dark-matter searches~\cite{Vasquez:2011}.
Furthermore, should SUSY or any other physics beyond the Standard Model (BSM) be discovered, independent 
studies of all three lepton flavours are necessary to investigate the coupling structure of the new physics, 
especially with regard to lepton universality. 
If squarks and gluinos are within the reach of the Large Hadron Collider (LHC), their production may be among the dominant SUSY processes.

This article reports on an inclusive search for squarks and gluinos produced via the strong interaction in events with jets, at least one hadronically decaying tau lepton, and large missing transverse momentum from undetected LSPs.
Two distinct topologies are studied, with one tau lepton (\onetau) or two or more tau leptons (\twotau) in the final state. These mutually exclusive channels are optimised separately.
The analysis is performed using $3.2~\ifb$ of proton--proton ($pp$) collision data at $\sqrt{s}=13 \tev$ recorded with the ATLAS detector at the LHC in 2015.  
Two SUSY scenarios are considered: a gauge-mediated SUSY-breaking (GMSB) model, and a simplified model of gluino pair production.
Previous searches based on the same final state have been reported by the ATLAS~\cite{SUSY-2012-07,SUSY-2013-10} and CMS~\cite{Chatrchyan:2013dsa} collaborations.

In GMSB models~\cite{Dine:1981gu,AlvarezGaume:1981wy,Nappi:1982hm}, SUSY breaking is communicated from a hidden sector to the visible sector by a set of messenger fields that share the gauge interactions of the SM.
SUSY is spontaneously broken in the messenger sector, leading to massive, non-degenerate messenger fields.
Gauginos acquire their masses via one-loop diagrams involving messengers. 
Squarks and sleptons, which do not couple directly to the messenger sector, get their masses from two-loop diagrams involving messengers and SM gauge bosons, or messengers and gauginos. 
The free parameters of GMSB models are the SUSY-breaking mass scale in the messenger sector ($\Lambda$), the messenger mass scale ($M_\mathrm{mes}$), 
the number of messenger multiplets ($N_5$) of the $\mathbf{5}+\mathbf{\bar 5}$ representation of $\text{SU}(5)$, the ratio of the two Higgs-doublet vacuum expectation values at the electroweak scale ($\tan\beta$), 
the sign of the Higgsino mass term in the superpotential ($\mathrm{ sign}(\mu)=\pm1$), and a gravitino-mass scale factor ($C_\mathrm{grav}$).
Masses of gauginos and sfermions are proportional to $\Lambda$, and scale as $N_5$ and $\sqrt{N_5}$, respectively.
The $M_\mathrm{mes}$ scale is required to be larger than $\Lambda$, to avoid tachyonic messengers and charge- and colour-breaking vacua, and lower than the Planck mass to suppress flavour violation.
The latter condition implies that the lightest supersymmetric particle is a very light gravitino.
The $C_\mathrm{grav}$ parameter, which results from the mechanism communicating SUSY breaking to the messengers, mainly affects the decay rate of the next-to-lightest supersymmetric particle (NLSP) into the LSP. 

As in previous ATLAS searches~\cite{SUSY-2012-07,SUSY-2013-10}, the GMSB model is probed as a function of $\Lambda$ and $\tan\beta$, and the other parameters are set to $M_\mathrm{mes} =250$~\tev, $N_{\mathrm 5}=3$, $\mathrm{sign}(\mu)=1$ and $C_\mathrm{grav}=1$.
For this choice of parameters, the NLSP is the lightest scalar tau ($\PSgti$) for large values of $\tan\beta$, while for lower $\tan\beta$ values, the $\PSgti$ and the superpartners of the right-handed electron and muon ($\PSe_{\text{R}},\PSgm_{\text{R}}$) are almost degenerate in mass.
The squark--antisquark production mechanism dominates at high values of $\Lambda$. A typical GMSB signal process is displayed in Fig.~\ref{fig:GMSBGraph}.
The value of $C_\mathrm{grav}$ corresponds to prompt decays of the NLSP.
The region of small $\Lambda$ and large $\tan\beta$ is unphysical since it leads to tachyonic states.

\begin{figure}[tb]
  \centering
  \subfigure[GMSB]{
    \includegraphics[width=.42\textwidth]{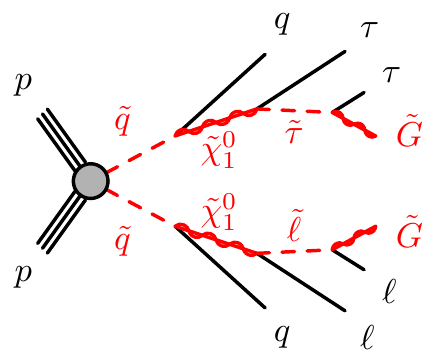}
    \label{fig:GMSBGraph}
  }
  \hspace{5mm}
  \subfigure[Simplified model]{
    \includegraphics[width=.42\textwidth]{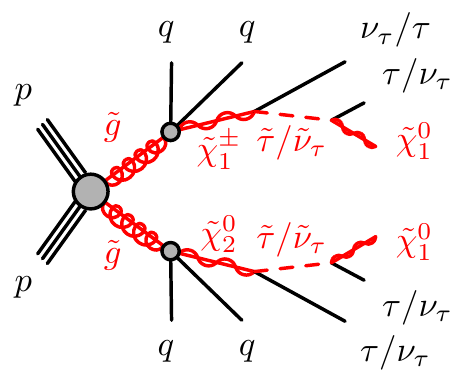}
    \label{fig:SimplifiedModelGraph}
  }
 \caption{Diagrams illustrating the two SUSY scenarios studied in this analysis. In the GMSB model, the scalar lepton $\tilde \ell$ is preferentially a scalar tau $\stau_1$ for high values of $\tan\beta$.}
  \label{fig:FeynmanDiagram}
\end{figure}%

The other signal model studied in this analysis is a simplified model of gluino pair production~\cite{SUSY-2014-06} in an $R$-parity-conserving scenario.
It is inspired by generic models such as the phenomenological MSSM~\cite{Djouadi:1998di,Berger:2008cq} with dominant gluino pair production, light $\tilde\tau_1$ and a $\tilde\chi_1^0$ LSP.
Gluinos are assumed to undergo a two-step cascade decay leading to tau-rich final states, as shown in Fig.~\ref{fig:SimplifiedModelGraph}. 
The two free parameters of the model are the masses of the gluino ($m_{\tilde g}$) and the LSP ($m_{\tilde\chi_1^0}$).
Assumptions are made about the masses of other sparticles, namely the $\tilde\tau_1$ and $\tilde\nu_\tau$ are mass-degenerate, and the $\tilde\chi_2^0$ and $\tilde\chi_1^\pm$ are also mass-degenerate, with
\begin{equation}
m_{\tilde\chi_1^\pm} = m_{\tilde\chi_2^0}= \frac{1}{2}(m_{\tilde g} + m_{\tilde\chi_1^0}) \; , \quad
m_{\tilde\tau_1}=m_{\tilde\nu_\tau}=\frac{1}{2}(m_{\tilde\chi_1^\pm} + m_{\tilde\chi_1^0}).
\end{equation}
Gluinos are assumed to decay to $\tilde{\chi}^{\pm}_{1} q \bar q'$ and $\tilde{\chi}^0_{2} q \bar q$ with equal branching ratios, where $q,q'$ denote generic first- and second-generation quarks.
Neutralinos $\tilde{\chi}^0_{2}$ are assumed to decay to $ \stau \tau$ and $ \tilde \nu_{\tau} \nu_{\tau}$ with equal probability, and charginos $\tilde{\chi}^{\pm}_{1}$ are assumed to decay to $ \tilde \nu_{\tau}\tau$ and $ \stau\nu_{\tau}$ with equal probability.
In the last step of the decay chain, $\stau $ and $\tilde \nu_{\tau}$ are assumed to decay to $\tau \tilde{\chi}^0_{1}$ and $ \nu_\tau \tilde{\chi}^0_{1}$, respectively. 
All other SUSY particles are kinematically decoupled. 
The topology of signal events depends on the mass splitting between the gluino and the LSP.  
The sparticle decay widths are assumed to be small compared to sparticle masses, such that they play no role in the kinematics.

\section{The ATLAS detector}
\label{sec:detector}
The ATLAS experiment is described in detail in Ref.~\cite{PERF-2007-01}.
It is a multi-purpose detector with a forward-backward symmetric cylindrical
geometry and a solid angle\footnote{ATLAS uses a right-handed coordinate
system with its origin at the nominal interaction point in the
centre of the detector and the $z$-axis along the beam pipe. The
$x$-axis points from the interaction point to the centre of the LHC ring and the
$y$-axis points upward. Cylindrical coordinates $(r,\phi)$ are used
in the transverse plane, $\phi$ being the azimuthal angle around the
beam pipe. The pseudorapidity is defined in terms of the polar angle
$\theta$ as $\eta=-\ln\tan(\theta/2)$.} coverage of nearly 4$\pi$.

The inner tracking detector (ID), covering the region $|\eta|<2.5$, consists
of a silicon pixel detector, a silicon microstrip detector and a
transition radiation tracker. 
The innermost layer of the pixel detector, the insertable B-layer~\cite{IBL}, was installed between Run 1 and Run 2 of the LHC. The inner detector is surrounded by a thin
superconducting solenoid providing a \unit[2]{T} magnetic field, and by
a finely segmented lead/liquid-argon (LAr) electromagnetic calorimeter covering the region $|\eta|<3.2$.
An steel/scintillator-tile hadronic calorimeter provides coverage in
the central region $|\eta|<1.7$. The end-cap and
forward regions, covering the pseudorapidity range $1.5<|\eta|<4.9$, are instrumented with electromagnetic and hadronic
LAr calorimeters, with either steel, copper or tungsten as the absorber material.

A muon spectrometer system incorporating large
superconducting toroidal air-core magnets surrounds the calorimeters. Three layers of precision wire chambers 
provide muon tracking coverage in the range $|\eta|<2.7$, while dedicated fast chambers are used for triggering in the region $|\eta|<2.4$.

The trigger system, composed of two stages, was upgraded~\cite{ATL-DAQ-PUB-2016-001} before Run 2. 
The Level-1 trigger system, implemented with custom hardware, uses information from calorimeters and muon chambers to reduce the event rate from 40~MHz to a maximum of 100~kHz.
The second stage, called the High-Level Trigger (HLT), reduces the data acquisition rate to about 1~kHz. 
The HLT is based on software and runs reconstruction algorithms similar to those used in the offline reconstruction.

\section{Data and simulation samples}
\label{sec:dataMC}
The data used in this analysis consist of $pp$ collisions at a centre-of-mass energy of $\sqrt{s}=13$~\tev\ delivered by the LHC with a 25~ns bunch spacing and recorded by the ATLAS detector from August to November 2015.
Data quality requirements are applied to ensure that all sub-detectors were operating normally, and that LHC beams were in stable-collision mode.
The integrated luminosity of the resulting data set is $3.16 \pm 0.07$~\ifb.

Simulated Monte Carlo (MC) event samples are used to model both the SUSY signals and SM backgrounds, except multi-jet production, which is evaluated from data.
All MC samples are generated at $\sqrt{s}=13$~\tev.
In addition to the hard-scattering process, soft $pp$ interactions (pile-up) are included in the simulation using the {\sc Pythia}~8.186~\cite{Sjostrand:2014zea} generator with the A2~\cite{ATL-PHYS-PUB-2012-003} set of tuned parameters (tune) and MSTW2008LO~\cite{Martin:2009iq} parton density function (PDF) set. Generated events are reweighted such that the average number of $pp$ interactions per bunch crossing has the same distribution in data and simulation.
For SM background samples, the interactions between generated particles and the detector material are simulated~\cite{SOFT-2010-01} using {\sc Geant4}~\cite{Agostinelli:2002hh} and a detailed description of the ATLAS detector.
In the case of signal samples, a parameterised fast simulation~\cite{ATL-SOFT-PUB-2014-01} is used to describe the energy deposits in the calorimeters.

The $W$+jets and $Z$+jets processes are simulated with the {\sc Sherpa}~2.1.1~\cite{Gleisberg:2008ta} generator.
Matrix elements (ME) are calculated for up to two partons at next-to-leading order (NLO) and up to four additional partons at leading order (LO) in perturbative QCD using the {\sc OpenLoops}~\cite{Cascioli:2011va} and {\sc Comix}~\cite{Gleisberg:2008fv} matrix element generators, respectively.
The polarisation of tau leptons in $W(\tau\nu)$+jets and $Z(\tau\tau)$+jets events is handled by the {\sc TauSpinner}~\cite{Czyczula:2012ny} program.
The phase-space merging between the {\sc Sherpa} parton shower (PS)~\cite{Schumann:2007mg} and matrix elements follows the ME+PS@NLO prescription~\cite{Hoeche:2012yf}. 
The CT10~\cite{Lai:2010vv} PDF set is used in conjunction with dedicated parton-shower tuning.  
Simulated samples are generated in bins of the transverse momentum (\pt) of the vector boson. 
The inclusive cross sections are normalised to a next-to-next-to-leading-order (NNLO) calculation~\cite{Catani:2009sm} in perturbative QCD based on the {\sc FEWZ} program~\cite{Anastasiou:2003ds}.

For the simulation of $t\bar t$ and single-top-quark production in the $Wt$- and $s$-channels, the {\sc Powheg-Box} v2~\cite{Alioli:2010xd} generator is used with the CT10 PDF set for the matrix elements calculation. 
Electroweak $t$-channel single-top-quark events are generated using the {\sc Powheg-Box} v1 generator. 
This generator uses the four-flavour scheme for the NLO matrix element calculation together with the fixed four-flavour CT10f4 PDF set. 
For all top quark processes, top quark spin correlations are taken into account (for $t$-channel production, top quarks are decayed using {\sc MadSpin}~\cite{Artoisenet:2012st}). 
The parton shower, hadronisation, and underlying event are simulated using {\sc Pythia}~6.428~\cite{Sjostrand:2006za} with the CTEQ6L1~\cite{Pumplin:2002vw} PDF set and the corresponding Perugia~2012 tune~\cite{Skands:2010ak}. 
Cross sections are calculated at NNLO in perturbative QCD with resummation of next-to-next-to-leading logarithmic (NNLL) soft gluon terms using the {\sc Top++} 2.0 program~\cite{Czakon:2011xx}.

Diboson production is simulated using the {\sc Sherpa} 2.1.1 generator with the CT10 PDF set. 
Processes with fully leptonic final states are calculated with up to one ($4\ell$, $2\ell+2\nu$) or no partons ($3\ell+1\nu$) at NLO and up to three additional partons at LO. 
Diboson processes with one of the bosons decaying hadronically and the other leptonically are simulated with up to 1 ($ZZ$) or 0 ($WW$, $WZ$) partons at NLO and up to 3 additional partons at LO. 
The generator cross sections are used for these samples.

The simplified-model signal samples are generated using {\sc MG5\_aMC@NLO} v2.2.3~\cite{Alwall:2014hca} interfaced to {\sc Pythia} 8.186 with the A14 tune~\cite{ATL-PHYS-PUB-2014-021} for the modelling of the parton shower, hadronisation and underlying event. The ME calculation is performed at tree level and includes the emission of up to two additional partons. The PDF set used for the generation is NNPDF23LO~\cite{Ball:2013hta}. The ME--PS matching is done using the CKKW-L prescription~\cite{Lonnblad:2001iq}, with a matching scale set to one quarter of the gluino mass. 
The GMSB signal samples are generated with the {\sc Herwig++}~2.7.1~\cite{Bahr:2008pv} generator, with CTEQ6L1 PDFs and the UE-EE-5-CTEQ6L1 tune~\cite{Gieseke:2012ft}, 
using input files generated  in the SLHA format with the {\sc SPheno} v3.1.12~\cite{Porod:2011nf} program.
The parton shower evolution is performed using an algorithm described in Refs.~\cite{Bahr:2008pv,Marchesini:1983bm,Marchesini:1987cf,Gieseke:2003rz}.
Signal cross sections are calculated at NLO in the strong coupling constant, adding the resummation of soft gluon emission at next-to-leading-logarithm accuracy~\cite{Beenakker:1996ch,Kulesza:2008jb,Kulesza:2009kq,Beenakker:2009ha,Beenakker:2011fu}. The nominal cross section and the uncertainty are taken from an envelope of cross-section predictions using different PDF sets and factorisation and renormalisation scales, as described in Ref.~\cite{Kramer:2012bx}.

\section{Reconstruction of final-state objects}
\label{sec:objects}
This analysis primarily requires the presence of jets, hadronically decaying tau leptons and missing transverse momentum in the final state.
Jet $b$-tagging is used to separate the top quark background from vector bosons produced in association with jets ($V$+jets, where $V=W,Z$).
Electrons and muons are vetoed in the \onetau channel, and muons are explicitly used in the \twotau channel for background modelling studies.

Primary vertices are reconstructed using inner-detector tracks with $\pt>400 \mev$ that satisfy requirements on the number of hits in silicon tracking devices~\cite{ATL-PHYS-PUB-2015-026}.
Primary vertex candidates are required to have at least two associated tracks, and the candidate with the largest
$\sum\pt^2$ is chosen as the primary vertex.

Jets are reconstructed with the anti-$k_t$ clustering algorithm~\cite{Cacciari:2008gp} with a distance parameter $R=0.4$.
Clusters of topologically connected calorimeter cells~\cite{Aad:2016upy} with energy above the noise threshold, calibrated at the electromagnetic energy scale, are used as input.
The jet energy is calibrated using a set of global sequential calibrations~\cite{ATLAS-CONF-2015-002,ATL-PHYS-PUB-2015-015}. 
Energy from pile-up interactions is subtracted based on the jet area and the median energy density computed for each event~\cite{PERF-2014-03}.
Jets are required to have $\pt > 20 \gev$ and $|\eta| < 2.8$.
The discrimination between hard-interaction jets and pile-up jets is achieved by a jet-vertex-tagging algorithm~\cite{ATLAS-CONF-2014-018}.
Events with jets originating from cosmic rays, beam background or detector noise are vetoed using the {\it loose} quality requirements defined in Ref.~\cite{ATLAS-CONF-2015-029}. 
Jets containing $b$-hadrons ($b$-jets) are identified using a multivariate algorithm exploiting the long lifetime, high decay
multiplicity, hard fragmentation, and large mass of $b$-hadrons~\cite{ATL-PHYS-PUB-2015-022}. The $b$-tagging algorithm identifies genuine $b$-jets
with an efficiency of approximately 70\% in simulated $t\bar t$ events.
The rejection rates for $c$-jets, hadronically decaying tau leptons, and light-quark or gluon jets are approximately 8, 26 and 440, respectively~\cite{ATL-PHYS-PUB-2015-039}.

Electron candidates are reconstructed from an isolated energy deposit in the electromagnetic calorimeter matched
to an inner-detector track. They are required to have $\pt > 10 \gev$, $|\eta| < 2.47$, and candidates reconstructed in the transition region between the barrel and end-cap calorimeters ($1.37<|\eta|<1.52$) are discarded.
Electrons are required to satisfy a {\it loose} likelihood identification requirement~\cite{PERF-2013-05,ATL-PHYS-PUB-2015-041} based on calorimeter shower shapes and track properties. 
The significance of the transverse impact parameter of the electron track is required to be less than five.

Muon candidates are reconstructed in the region $|\eta| < 2.5$ from muon spectrometer tracks matching
ID tracks. Muons are required to have $\pt > 10 \gev$ and pass the {\it medium} identification requirements
defined in Ref.~\cite{Aad:2016jkr}, based on the number of hits in the ID and muon spectrometer, and the compatibility of the charge-to-momentum ratio measured in the two detector systems.

Hadronically decaying tau leptons are reconstructed~\cite{ATL-PHYS-PUB-2015-045} from anti-$k_t$ jets with $E_\mathrm{T}\geq10\gev$ and $|\eta|<2.5$ calibrated with a local cluster weighting technique~\cite{PERF-2011-03}.
Tau candidates are built from clusters of calorimeter cells within a cone of size $\Delta R \equiv \sqrt{\smash[b]{(\Delta\eta)^2+(\Delta\phi)^2}}=0.2$ centred on the jet axis.
Tau energy scale corrections are applied to subtract the energy originating from pile-up interactions and correct for the calorimeter response. 
Tau leptons are required to have $\pt>20 \gev$, and candidates reconstructed within the transition region $1.37 < |\eta| <1.52$ are discarded.
Tau leptons are required to have either one or three associated tracks, with a charge sum of $\pm 1$.
A boosted-decision-tree discriminant  is used to separate jets from tau leptons.
It relies on track variables from the inner detector as well as longitudinal and transverse shower-shape variables from the calorimeters.
The analysis makes use of {\it loose} and {\it medium} tau leptons, corresponding to identification working points with efficiencies of 60\% and 55\% for one-track tau leptons, respectively, and 50\% and 40\% for three-track tau leptons, respectively.
Electrons mis-identified as 1-track tau leptons are rejected by imposing a \pt- and $|\eta|$-dependent requirement on the electron likelihood, which provides a constant efficiency of 95\% for real tau leptons, with an inverse background efficiency (rejection factor) for electrons ranging from 30 to 150 depending on the $|\eta|$ region.

The missing transverse momentum vector \ptmissv, whose magnitude is denoted by \met, is defined as the negative vector sum of the transverse momenta of all identified physics objects (electrons, muons, jets, and tau leptons) and an additional soft-term. 
Therefore, the \met calculation benefits from the dedicated calibration for each final-state object.
The soft-term is constructed from all the tracks with $\pt>400\mev$ originating from the primary vertex which are not associated with any physics object. 
This track-based definition makes the soft-term largely insensitive to pile-up~\cite{ATL-PHYS-PUB-2015-027}.

After object reconstruction, an {\it overlap-removal} procedure is applied to remove ambiguities in case the same object is reconstructed by several algorithms.
Tau candidates are discarded if they are found to overlap with a light lepton (electron or muon). If an electron and a muon are reconstructed using the same inner-detector track, the electron is discarded. 
For overlapping jets and electrons, the electron is kept. Light leptons in the vicinity of a jet are considered to originate from secondary decays within the jet and are discarded. 
Finally, in case a jet is reconstructed as a tau lepton, the tau candidate is retained.
The successive steps of this procedure are summarised in Table~\ref{tab:OverlapRemoval}.
The final-state objects considered in the analysis are those surviving the overlap-removal algorithm. 

\begin{table}[tb]
  \centering
 \caption{Overview of the successive steps in the overlap removal algorithm.}
  \begin{tabular}{clll}
  \hline\hline
         & Object discarded            & Object kept      & Matching condition                  \\
    \hline
    1.   & loose tau                 &  electron            & $\Delta R < 0.2$          \\
    2.   & loose tau                 &  muon                & $\Delta R < 0.2$           \\
    3.   &  electron         &  muon                & shared inner-detector track    \\ 
    4.   & jet                       &  electron            & $\Delta R < 0.2$        \\
    5.   &  electron         & jet                          & $\Delta R < 0.4$        \\
    6.   &  muon             & jet                          & $\Delta R < 0.4$        \\
    7.   & jet                       & loose tau                    & $\Delta R < 0.2$        \\
  \hline\hline
  \end{tabular}
   \label{tab:OverlapRemoval}
\end{table}

\section{Event selection}
\label{sec:selection}

The trigger used in this analysis is the lowest-threshold missing transverse momentum trigger that was active without restrictions during the whole 2015 data-taking period.
The efficiency of that trigger is measured using data collected by a set of single-jet triggers, with events containing at least two jets and one loose tau lepton. 
The trigger is found to have an efficiency greater than 99\% when requiring $\met > 180 \gev$ and a jet with $\pt > 120 \gev$ in the offline selection, which is referred to as {\it trigger plateau} conditions.

After trigger requirements, a pre-selection common to both channels is applied to ensure that only well-reconstructed events enter the analysis.
Events containing no reconstructed primary vertex, mis-measured muon tracks, cosmic muon candidates, jets originating from calorimeter noise or reconstructed near inactive areas in the calorimeter, are vetoed. 
The presence of a second jet with $\pt > 20 \gev$ is required.
To suppress the contribution from multi-jet events where large \met would arise from jet energy mis-measurement, a minimum angular separation in the transverse plane is imposed between either of the two leading jets and the missing transverse momentum, $\Delta\phi(\text{jet}_{1,2},\ptmissv)>0.4$.

As part of the pre-selection in the \onetau channel, events are required to contain exactly one medium tau lepton and no light lepton.
The veto against electrons and muons was used in Run 1 to ensure the \onetau channel did not overlap with the $e\tau$ and $\mu\tau$ channels, which are not part of the present analysis.
The lepton veto does not affect the expected sensitivity to the simplified model, and has been kept in the \onetau channel.
Events with two or more loose tau leptons are rejected to make the \onetau and \twotau channels statistically independent.
In the \twotau channel, at least two loose tau leptons are required at pre-selection level.
No veto is applied against light leptons, as this would cause a sizeable selection inefficiency for GMSB signals.

In each channel, signal regions (SRs) are defined for several signal scenarios. 
The following kinematic variables are found to provide discrimination between signal and background, or between backgrounds themselves:
\begin{itemize}
\renewcommand\labelitemi{--}
\item{the transverse mass of the system formed by \ptmissv and a lepton $\ell$ assumed to be massless,
  \begin{equation}
  \mTl \equiv m_\mathrm{T}(\ell,\ptmissv)=\sqrt{2\pt^{\ell}\met(1-\cos\Delta\phi(\ell,\ptmissv))}~.
  \end{equation} 
  For events where the missing transverse momentum and the lepton originate from a $W\to\ell\nu$ decay, the $\mTl$ distribution exhibits a 
  Jacobian peak at
  the $W$ boson mass.
  In the \onetau channel, the $\mT^\tau$ variable is used. In the \twotau channel, the \mTTT variable based on the two leading tau leptons is used, and $m_\mathrm{T}^\mu$ is used for specific selections requiring a muon;}
\item{the scalar sum of the transverse momenta of all tau leptons and jets in the event, \\
    $\HT=\sum\limits_i \pt^{\mathrm{\tau}_i}+ \sum\limits_j \pt^{\mathrm{jet}_j}$~;}
\item{the magnitude of the missing transverse momentum, \met;}
\item{the effective mass, $\meff=\HT+\met$;}
\item{the $\mttwo^{\tau\tau}$ variable~\cite{Lester:1999tx,Lester:2014yga}, also called {\it stransverse mass}, computed as
\begin{equation}
\mttwo^{\tau\tau} =\sqrt{ \min\limits_{{\vec{p}_\mathrm{T}^{\,a}+ \vec{p}_\mathrm{T}^{\, b} = \ptmissv }}\left(\max\left[m_\mathrm{T}^2(\tau_1,\vec{p}_\mathrm{T}^{~a}) , m_\mathrm{T}^2(\tau_2,\vec{p}_\mathrm{T}^{~b})   \right]\right)}~,
\label{eq:mttwo}
\end{equation}
where $(a,b)$ refers to two invisible particles that are assumed to be produced with transverse momentum vectors $\vec{p}_\mathrm{T}^{~a,b} $. In this calculation, $(a,b)$ are assumed to be massless.
The $\mttwo^{\tau\tau}$ distribution has a kinematic endpoint for processes where massive particles are pair-produced, each particle decaying to a tau lepton and an undetected particle.
In cases where multiple tau leptons are produced in a decay chain, there is no a-priori way to select the pair leading to the desired characteristic.
For events with more than two tau-lepton candidates, $\mttwo^{\tau\tau}$ is hence calculated using all possible tau-lepton pairs and the largest value is chosen;}
 \item{the sum of the transverse masses of all jets and of the two leading tau leptons in the event,\\ 
 $\mT^\mathrm{sum} = \mTTT+ \sum\limits_i m_{\mathrm T}^{\mathrm {jet}_i}$~,\\
where $m_{\mathrm T}^{\mathrm {jet}}$ is defined analogously to $\mTl$ ; }
\item{the total number of jets, \njet;}
\item{the number of $b$-tagged jets, \nbjet.}
\end{itemize}
These variables are also used for the selection of control regions (CRs) and validation regions (VRs) in the context of background modelling studies, as discussed in Section~\ref{sec:backgrounds}.

Figure~\ref{fig:preselection} shows example kinematic distributions at pre-selection level. The dominant backgrounds are $W(\tau\nu)$+jets and $t\bar t$ production in the \onetau channel, with subdominant contributions from $Z(\nu\nu)$+jets and $Z(\tau\tau)$+jets. 
In the \twotau channel, the pre-selection is dominated by $t\bar t$, $W(\tau\nu)$+jets, and $Z(\tau\tau)$+jets events. The multi-jet and diboson backgrounds do not contribute significantly.

\begin{figure}[htb]
 \centering
  \subfigure[$H_\mathrm{T}$ distribution in the \onetau channel]{\includegraphics[width=.48\textwidth]%
 		{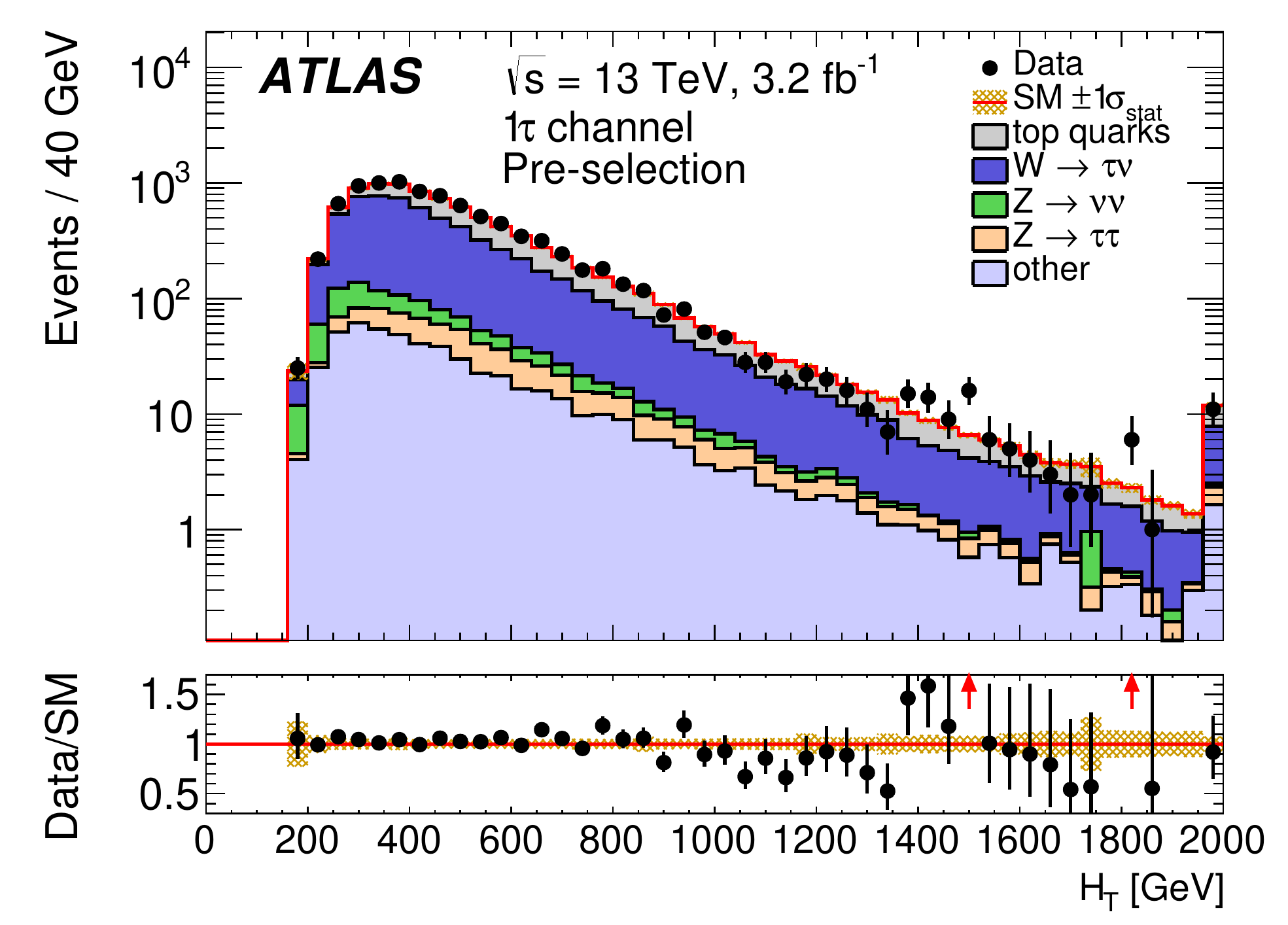}
                 }
   \subfigure[$m_\mathrm{T}^{\tau_1}+m_\mathrm{T}^{\tau_2}$ distribution in the \twotau channel]{\includegraphics[width=.48\textwidth]%
                 {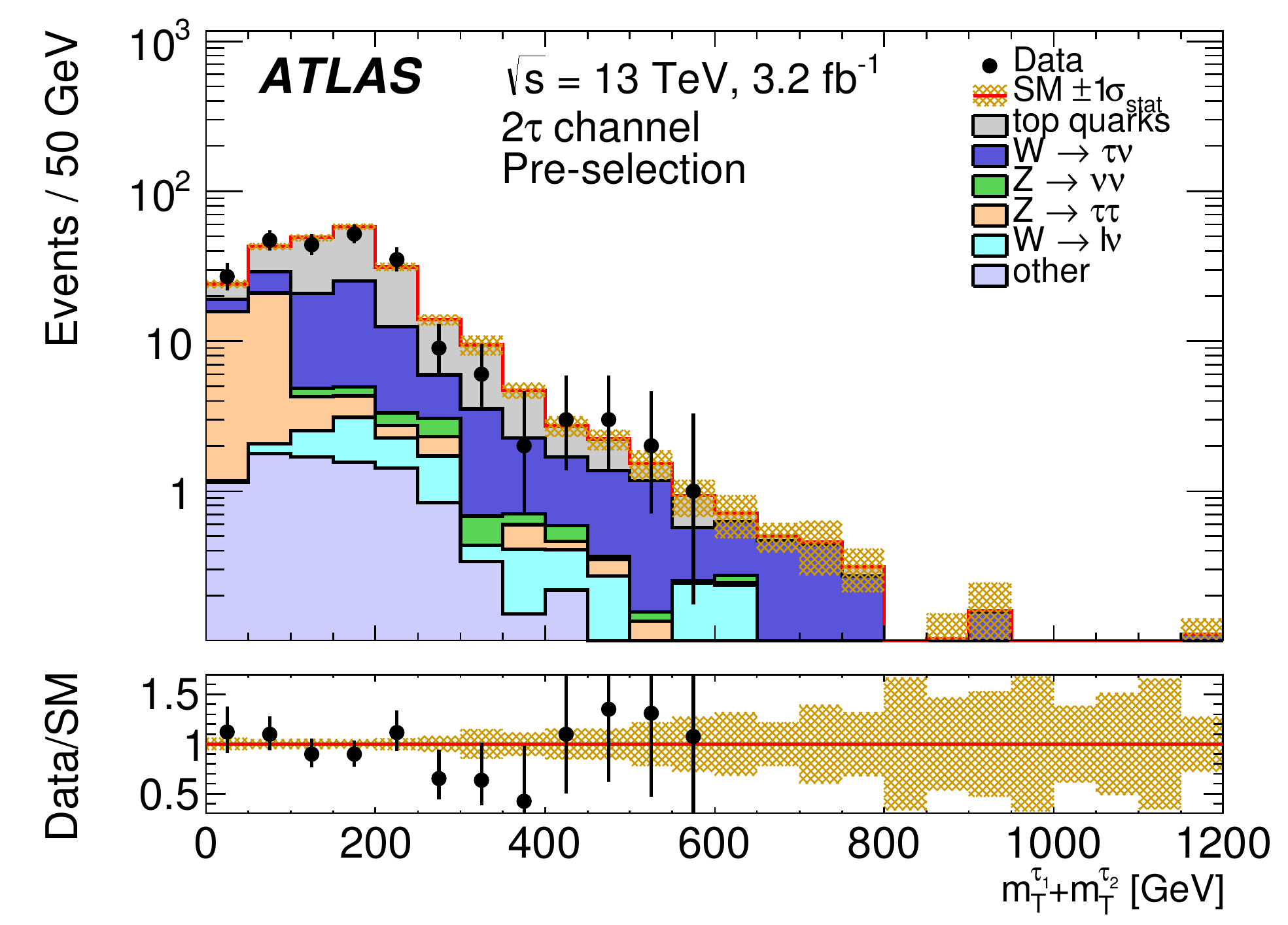}
                 } 
  \caption{Kinematic distributions at pre-selection level, for the \onetau and \twotau channels. 
  The last bin includes overflow events. The shaded bands indicate the statistical uncertainties in the background predictions.
  The contribution labelled as ``other'' includes diboson and multi-jet events, and the $V$+jets processes not explicitly listed in the legend.
  Red arrows in the Data/SM ratio indicate bins where the corresponding entry falls outside the plotted range.}
   \label{fig:preselection}
\end{figure}

In the \onetau channel, three characteristic regions in the $(m_{\tilde g},m_{\tilde\chi_1^0})$ parameter space of the simplified model are chosen as benchmark scenarios for the SR optimisation, with small ($<100\gev$), medium ($500$--$900\gev$) and large ($>1200 \gev$) mass splittings between the gluino and the LSP.
In the following, the associated SRs are called Compressed, Medium-Mass and High-Mass SRs, respectively.
The Compressed SR exploits topologies where a high-\pt jet from initial-state radiation (ISR) recoils against the pair of gluinos. 
In this situation, the soft visible particles produced in the gluino decay receive a transverse Lorentz boost.
While tau leptons and jets from gluino decays typically have low \pt, such events have substantial \met since both LSPs tend to be emitted oppositely to the ISR jet in the transverse plane. 
In the case of large mass splitting, high-\pt jets come mainly from gluino decays, and a high-\pt requirement on the first two leading jets is effective in rejecting background without inducing a large inefficiency for the signal.
A requirement on the transverse mass $\mT^\tau$ is applied in the Medium-Mass and High-Mass SRs to suppress $W(\tau\nu)$+jets events as well as semileptonic $t\bar t$ events with a tau lepton in the final state.
The \HT variable is also used in these two SRs, as \HT increases for signal events with increasing mass splittings.
No SR is defined for the GMSB model, as the expected sensitivity in the \onetau channel is significantly lower than that in the \twotau channel.

In the \twotau channel, two SRs are defined for the simplified model to cover small ($<900\gev$) and large ($>1200 \gev$) mass-splitting scenarios.
The Compressed SR imposes a requirement on $\mttwo^{\tau\tau}$ to exploit the fact that most of SM background contributions exhibit a kinematic endpoint around the $W$ or $Z$ boson mass, which is not the case for tau leptons produced in the cascade decay of gluinos.
A requirement is also applied on \mTSum to take advantage of the large \met and the large jet and tau lepton multiplicity that is expected for signal events.
The High-Mass SR includes a requirement on \HT, which is efficient for high-mass gluino signals.
A requirement on \mTTT is also applied. 
In $Z(\tau\tau)$+jets events, where \met originates from neutrinos from tau lepton decays, 
the trigger plateau requirement selects events with a high-\pt $Z$ boson recoiling against jets in the transverse plane.
This topology leads to tau leptons with a small $\Delta\phi$ separation, which results in low values of \mTTT given that tau neutrinos are themselves collimated with the visible decay products of tau leptons.
For dileptonic $t\bar t$ events with two tau leptons and large genuine \met, and for $W(\tau\nu)$+jets events and semileptonic $t\bar t$ events with a high-\pt jet mis-identified as a tau lepton, \mTTT can be larger, 
but even larger values are expected for high-mass gluino signals.
A signal region is also defined for the GMSB model, and targets more specifically squark--antisquark production rather than gluino pair production in the region $\Lambda \gtrsim 80 \tev$, not excluded by Run 1 searches.  
Among the distinctive features that give large discrimination power to \HT, decay chains are potentially longer than in the simplified model, and the almost-massless gravitino LSP leaves more phase space to other particles in the decay.
Table~\ref{tab:sroverview} summarises the selection criteria for all the SRs of the \onetau and \twotau channels.

\begin{table}
  \centering
\caption{Selection criteria for the signal regions (SRs) of the \onetau channel (top) and the \twotau channel (bottom).}

{\renewcommand\arraystretch{1.2} 

 \begin{tabular}{l|c|c|c}
  \hline
  \hline

    \onetau channel                      &Compressed SR & Medium-Mass SR & High-Mass SR \\
  \hline
Trigger plateau &\multicolumn{3}{c}{$\met> 180 \gev$, $\pt^\mathrm{jet_1} >  120 \gev$}\\ 
                \hline
Tau leptons            &\multicolumn{3}{c}{$N_\tau^\text{loose} = N_\tau^\text{medium}=1$, $\pt^\tau>20 \gev$} \\ 
        \hline
Light leptons   &\multicolumn{3}{c}{$N_\ell=0\qquad$}\\
                \hline 
Multi-jet rejection      &\multicolumn{3}{c}{$\Delta\phi(\text{jet}_{1,2},\ptmissv)\geq 0.4$}\\
            \hline
\hline
\tauPt&$< 45 \gev$&--&-- \\
$\pt^\mathrm{jet_1}$ &$>  300 \gev$&--&$>  220 \gev$ \\
$\pt^\mathrm{jet_2}$& -- &--&$> 220 \gev$ \\
\njet                   & $\geq 2$&$\geq5$&$\geq5$ \\
$\mT^\tau$&$>  80 \gev$&$>  200 \gev$&$>  200 \gev$ \\
\met                    &$> 300 \gev$&$> 300 \gev$&-- \\
\HT                     &--&$> 550 \gev$&$> 550 \gev$ \\
\hline
\hline
 \end{tabular}

\vspace{20pt}

 \begin{tabular}{l|c|c|c}
  \hline
  \hline
 \twotau channel                        &Compressed SR & High-Mass SR & GMSB SR\\
  \hline
Trigger plateau &\multicolumn{3}{c}{$\met> 180 \gev$, $\pt^\mathrm{jet_1} > 120 \gev$}\\ 
                \hline
Tau leptons            &\multicolumn{3}{c}{$N_\tau^\text{loose}\geq2$, $\pt^\tau>20 \gev$}\\ 
        \hline
Multi-jet rejection      & \multicolumn{3}{c}{$\Delta\phi(\text{jet}_{1,2},\ptmissv)\geq 0.4$} \\
                        
            \hline
\hline
\mTTT	&--	&$> 350 \gev$&$>  150 \gev$\\
\HT                     &--& $>  800 \gev$&$>  1700 \gev$\\
\njet                   &$\geq2$&$\geq3$&$\geq2$\\
$\mttwo^{\tau\tau}$                  &$> 60 \gev$ & --&--\\
$\mT^\mathrm{sum}  $     &$> 1400 \gev$&-- &--\\
\hline
\hline
 \end{tabular}
}
\label{tab:sroverview}
\end{table}

\section{Background estimation}
\label{sec:backgrounds}

To predict the background contributions in the SRs, the normalisation of the dominant backgrounds is fitted to data in dedicated CRs.
In each channel and for each SR, a simultaneous fit over the relevant CRs is performed using {\sc HistFitter}~\cite{Baak:2014wma} to extract these normalisation factors.
Control regions are designed to have an enhanced contribution from a single background process, with contributions from other backgrounds as small as possible to reduce the uncertainties originating from correlations between CRs.
Furthermore, CRs are defined in phase-space regions close to that of SRs, to avoid the extrapolation of the background normalisations over very different kinematic regimes.

A set of VRs is defined in intermediate phase-space regions between a SR and its associated CRs, where signal contributions are small. 
These VRs are not part of the fit; they are used to compare the fitted background predictions with the observed data in the vicinity of the SRs before unblinding those.

\subsection{Vector-boson and top quark backgrounds}

In both channels, the dominant backgrounds originate from SM processes involving the top quark or a massive vector boson and jets. 
These two backgrounds can be separated from each other by either requiring or vetoing the presence of a $b$-tagged jet in the event.
In addition, a tau-lepton candidate can be either a genuine tau lepton ({\it true tau lepton}) from a $W(\tau\nu)$ decay or a jet mis-identified as a tau lepton ({\it fake tau lepton}), which leads to two types of CRs.

In CRs targeting true tau-lepton contributions, the normalisation factor is used to absorb the theoretical uncertainties in cross-section computations, the experimental uncertainties in the integrated luminosity, and potential differences in the tau-lepton reconstruction and identification efficiencies between data and simulation.
In the case of fake tau-lepton contributions, the normalisation factor combines several effects: the quark/gluon composition of jets mis-identified as tau leptons, which is process-dependent, the parton shower and hadronisation models of the generator, and 
the modelling in the simulation of jet shower shapes in the calorimeter, which mainly depends on the {\sc Geant4} hadronic interaction model and the modelling of the ATLAS detector.
Other contributions affecting the background normalisation include the modelling of the kinematics and acceptance of background processes.
These contributions are absorbed into the true and fake tau-lepton normalisation factors in the \onetau channel, while they are treated as a separate {\it kinematic} normalisation factor in the \twotau channel to avoid double-counting 
(true- or fake-tau normalisation factors are applied to each tau lepton, while the kinematic normalisation factor is applied once per event). 

In the \onetau channel, four CRs are defined, with four associated normalisation factors.
They target the top quark background (including $t\bar t$ and single-top-quark processes) with either a true or a fake tau lepton, 
and $V$+jets events with either a true or a fake tau lepton, respectively dominated by $W(\tau\nu)$+jets and $Z(\nu\nu)$+jets processes.
The discrimination between true and fake tau-lepton contributions is achieved by a requirement on $m_\mathrm{T}^\tau$.
A common set of four CRs is defined for the Medium-Mass and High-Mass SRs, due to the similarity of background compositions and event kinematics. These are separated from the SRs by an upper bound on $\mT^\tau$.
Another set of four CRs is defined for the Compressed SR, to study more specifically the background modelling for low-\pt tau leptons. These CRs are separated from the SR by an upper bound on \met.
The selection criteria defining these CRs are summarised in Tables~\ref{tab:onetau:CRVR_Comp} and \ref{tab:onetau:CRVR_MH}. 
Figure~\ref{fig:1tau:CRplots} illustrates the good modelling of the background in the various CRs after the fit.

Three types of VRs are used in the \onetau channel for the Medium- and High-Mass SRs, to validate the background extrapolation from low-\HT to high-\HT for selections based on true tau leptons, 
and the extrapolations along \HT and $\mT^\tau$ for selections based on fake tau leptons. The separation of the VRs from the SRs is achieved by inverting the selections on $\mT^\tau$ or \HT.
For the Compressed SR, four VRs are used to validate the extrapolation of $V$+jets and top quark background predictions along \met, for both the true and fake tau-lepton selections. 
The separation of the VRs from the SRs is achieved by an inverted requirement on $\mT^\tau$ and \met, for true and fake tau-lepton VRs, respectively.
The selection criteria defining all the VRs in the \onetau channel are listed in Tables~\ref{tab:onetau:CRVR_Comp} and \ref{tab:onetau:CRVR_MH}. 

\begin{table}
    \caption{Selection criteria defining the control and validation regions of the Compressed SR in the \onetau channel. The pre-selection criteria are also applied although not shown in the table.    
	Unless mentioned, criteria apply to both the top quark and $V$+jets background regions. \protect\hypertarget{top}{Symbols}~$^\mathrm{\textcolor{blue}{*}}$ indicate that criteria are only applied to regions targeting the top quark background, and
	\protect\hypertarget{Vjets}{symbols}~$^\mathrm{\textcolor{blue}{\dag}}$ denote criteria only applied to $V$+jets background regions.
    }
\small{
{\setlength{\extrarowheight}{0.1cm}
  \begin{tabular}{l|cc|cc}
 \hline\hline
     \onetau channel                                 & True-tau CR                          & True-tau VR                  & Fake-tau CR                                     & Fake-tau VR                                                                           \\
\hline
    $\met$                            & $180<\met <300 \gev$ & $>300 \gev$ & $180<\met <250 \gev$           & $250<\met <300 \gev$                                                 \\
    $m_{\mathrm T}^\tau$                 & $<100 \gev$                           & $<80 \gev$                   & $100<m_{\mathrm T}^\tau<330 \gev$                    & $100<m_{\mathrm T}^\tau<330 \gev$                                                          \\
    $H_{\mathrm T}$                   & --                                     & $>400 \gev$                  & $<550^{\hyperlink{top}{*}} / 400 ^{\hyperlink{Vjets}{\dag}} \gev$                   & $>550^{\hyperlink{top}{*}} / 400^{\hyperlink{Vjets}{\dag}} \gev$   \\
    $\pt^\tau $                       & $<45 \gev$                            & $< 60\gev$                   & $< 45\gev$                                      & $< 60\gev$                                                                            \\
    $N_{\mathrm {jet}}$               & --                                     & $\geq 4^{\hyperlink{top}{*}}$              &  --                                   & $\geq 4^{\hyperlink{top}{*}}$                                                          \\
    $N_{b-\mathrm {jet}}$      & $\geq 1^{\hyperlink{top}{*}} / = 0^{\hyperlink{Vjets}{\dag}}$   & $\geq 1^{\hyperlink{top}{*}} / = 0^{\hyperlink{Vjets}{\dag}}$   &$\geq 1^{\hyperlink{top}{*}} / = 0^{\hyperlink{Vjets}{\dag}}$    & $\geq 1^{\hyperlink{top}{*}} / = 0^{\hyperlink{Vjets}{\dag}}$                                                          \\
    $\Delta\phi(\mathrm{jet_1},\ptmissv)$ & $>1.8^{\hyperlink{Vjets}{\dag}}$                   & --                            & $>2.0 ^{\hyperlink{top}{*}}$                                    & --                                                                                     \\
    $\Delta\phi(\tau,\ptmissv)$            & --                                     & --                            & $>1.0$ &              $>1.0$                                                                                                 \\
    $\met/\meff$                      & --                                     & --                            &$>0.2^{\hyperlink{top}{*}} / 0.3^{\hyperlink{Vjets}{\dag}}$    & $>0.2^{\hyperlink{top}{*}} / 0.3^{\hyperlink{Vjets}{\dag}}$                                                                                     \\
\hline\hline

    \end{tabular}
    }}
    \label{tab:onetau:CRVR_Comp}
\end{table}

\begin{table}
  \caption{Selection criteria defining the control and validation regions of the Medium-Mass and High-Mass SRs in the \onetau channel. The pre-selection criteria are also applied although not shown in the table.    
	Unless mentioned, criteria apply to both the top quark and $V$+jets background regions. \protect\hypertarget{top2}{Symbols}~$^\mathrm{\textcolor{blue}{*}}$ indicate that criteria are only applied to regions targeting the top quark background, and
	\protect\hypertarget{Vjets2}{symbols}~$^\mathrm{\textcolor{blue}{\dag}}$ denote criteria only applied to $V$+jets background regions.}
  \footnotesize{
  {\setlength{\extrarowheight}{0.1cm}
  \begin{tabular}{l|cc|ccc}
 \hline\hline
            \onetau channel                           & True-tau CR                           & True-tau VR                  & Fake-tau CR                                     & Fake-tau VR1                          & Fake-tau VR2                                  \\
  \hline
         $\met  $                            & --                                     & --                            & --                                               & $ < 400^{\hyperlink{Vjets2}{\dag}} \gev$         & --                                             \\ 
    $m_{\mathrm T}$                   & $<100 \gev$                           & $<100 \gev$                  & $100<m_{\mathrm T}<200 \gev$   & $100<m_{\mathrm T}<200 \gev$          & $200<m_{\mathrm T}<330 \gev$ \\
    $H_{\mathrm T}$                   & $<550 \gev$          & $>550 \gev$ & $<550^{\hyperlink{top2}{*}}/ 400^{\hyperlink{Vjets2}{\dag}} \gev$ & $>550^{\hyperlink{top2}{*}} \gev$      & $<550 \gev$                                   \\
		& -- &--  & -- 										&  $400 < H_\mathrm{T}< 700^{\hyperlink{Vjets2}{\dag}} \gev$ & --\\
    $\pt^\tau $                       & --                                     & $> 45\gev$                   & --                                               & --                                     & --                                             \\
    $N_{\mathrm {jet}}$               & $\geq 4$                              & $\geq 4$                     &    --                                 & $\geq 4^{\hyperlink{top2}{*}}$           & $\geq 4^{\hyperlink{top2}{*}}$              \\
    $N_{b-\mathrm {jet}}$      & $\geq 1^{\hyperlink{top2}{*}} / = 0^{\hyperlink{Vjets2}{\dag}}$   & $\geq 1^{\hyperlink{top2}{*}} / = 0^{\hyperlink{Vjets2}{\dag}}$   &$\geq 1^{\hyperlink{top2}{*}} / = 0^{\hyperlink{Vjets2}{\dag}}$    & $\geq 1^{\hyperlink{top2}{*}} / = 0^{\hyperlink{Vjets2}{\dag}}$   &      $\geq 1^{\hyperlink{top2}{*}} / = 0^{\hyperlink{Vjets2}{\dag}}$                                                  \\
     $\pt^\mathrm{jet_1} $                                 & --                                     & --                            & --                                               & $< 500^{\hyperlink{Vjets2}{\dag}} \gev$    & --                                             \\
    $\Delta\phi(\mathrm{jet_1},\ptmissv)$ & --                                     & --                            & $>2.0$                                          & --                                     & --                                             \\
    $\Delta\phi(\tau,\ptmissv)$           & --                                     & --                            & $>1.0$ &          $>1.0$ & $>1.0$                                                                                                      \\
    $\met/\meff$                      & --                                     & --       & $>0.2^{\hyperlink{top2}{*}} / 0.3^{\hyperlink{Vjets2}{\dag}}$ &$>0.2^{\hyperlink{top2}{*}} / 0.3^{\hyperlink{Vjets2}{\dag}}$ &$>0.2^{\hyperlink{top2}{*}} / 0.3^{\hyperlink{Vjets2}{\dag}}$ \\
 \hline\hline
  \end{tabular}
  }}
  \label{tab:onetau:CRVR_MH}
\end{table}

\begin{figure}[htb]
 \centering
  \subfigure[top true-tau CR of the Medium- and High-Mass SRs]{\includegraphics[width=.48\textwidth]%
		{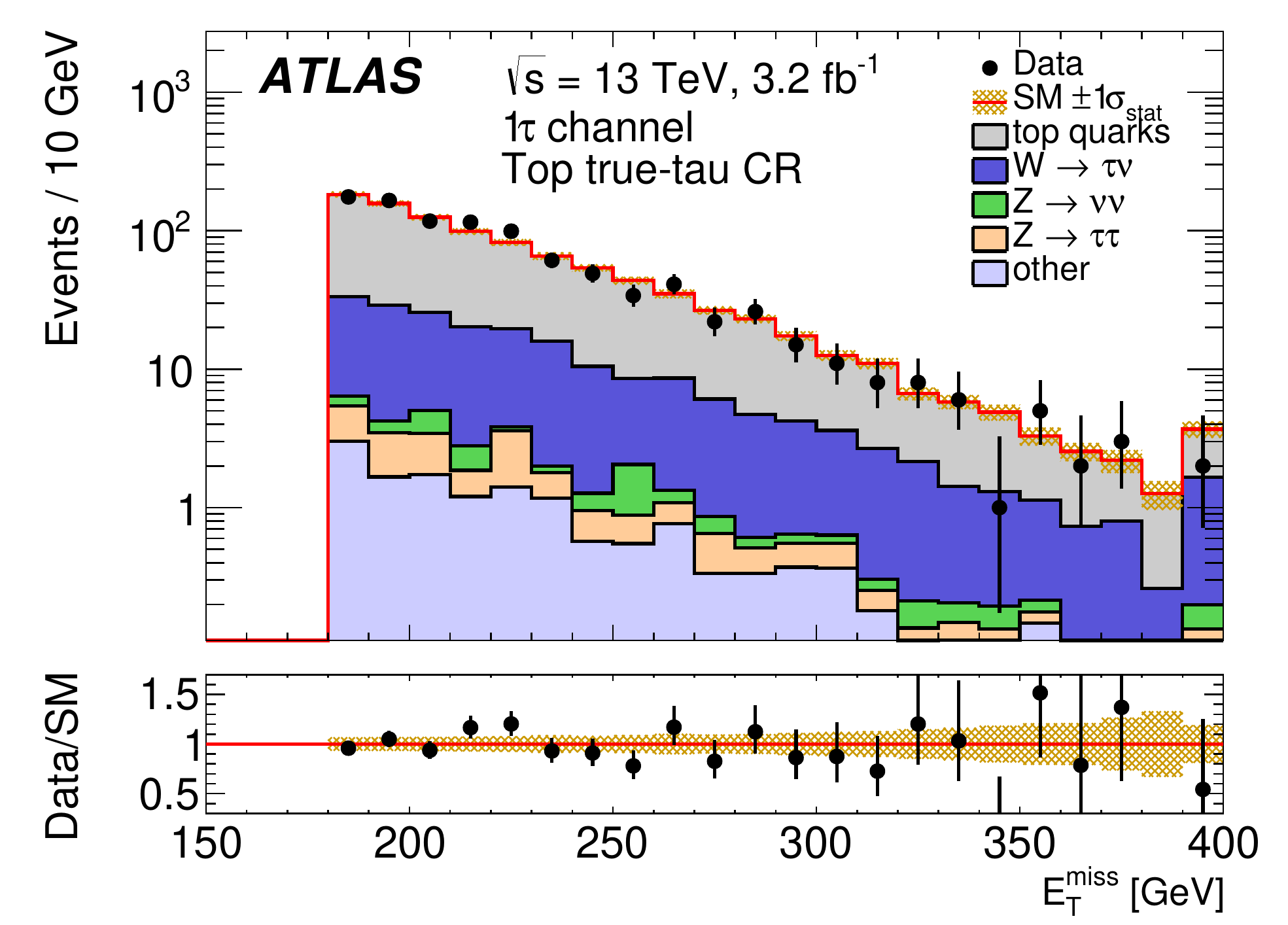}
                 }
  \subfigure[$V$+jets true-tau CR of the Compressed SR]{\includegraphics[width=.48\textwidth]%
		{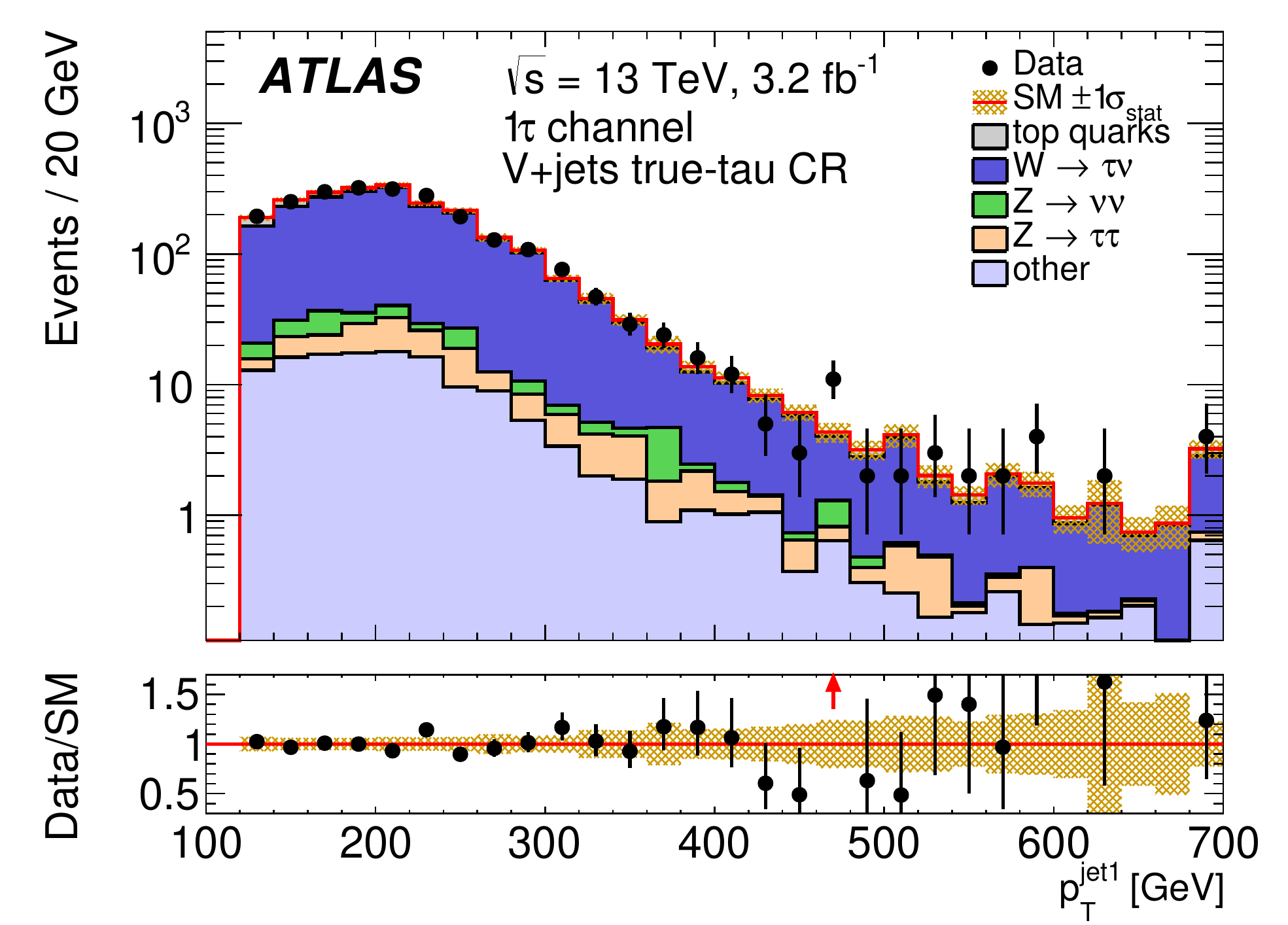}
                 }
  \caption{Missing transverse momentum and leading-jet \pt distributions in two different control regions of the \onetau channel after the fit, illustrating the overall background modelling in the CRs. 
  By construction, the total fitted background is equal to the number of observed events in each CR.
  The last bin includes overflow events. The shaded bands indicate the statistical uncertainties in the background predictions.
    Red arrows in the Data/SM ratio indicate bins where the corresponding entry falls outside the plotted range.}
   \label{fig:1tau:CRplots}
\end{figure}

In the \twotau channel, the dominant backgrounds are $Z(\tau\tau)$+jets and dileptonic top quark contributions (including $t\bar t $ and single-top quark processes) with two true tau leptons, $W(\tau\nu)$+jets and semileptonic top quark contributions with one true and one fake tau lepton, and $W(\ell\nu)$+jets and top quark contributions with two fake tau leptons.
Control regions are defined for $W$+jets and top quark backgrounds to extract normalisation factors related to the modelling of the process kinematics, and the modelling of real and fake tau leptons in the simulation. 
These CRs are separated from the SRs by replacing the requirement on the two tau leptons with requirements on different final-state objects, which stand in for true or fake tau leptons. 
To be independent from tau-lepton considerations, kinematic CRs are based on events with one muon, jets, \met, and without or with $b$-jets, to select $W(\mu\nu)$+jets and semileptonic top quark events with a final-state muon, respectively.
The fake tau-lepton CRs use the same baseline selections as the kinematic CRs, but in addition, the presence of a loose tau-lepton candidate is required.
Events with large $m_\mathrm{T}^\mu$ values are discarded to suppress the dileptonic top quark background with a muon and a true tau lepton.
The true tau-lepton CRs, which target $W(\tau\nu)$+jets and semileptonic top quark processes with a true tau lepton, are based on events with a loose tau lepton, jets and \met, without or with $b$-jets.
Similarly, contributions from fake tau leptons are suppressed by a requirement on $m_\mathrm{T}^\tau$.
A separate CR is designed to study $Z(\tau\tau)$+jets events by inverting the \mTTT and \HT requirements from the SRs. 
This selection requires two loose tau leptons of opposite electric charge.
The selection criteria defining the various CRs are summarised in Table~\ref{tab:twotau:CRs}. 
Figure~\ref{fig:2tau:CRplots} illustrates the good background modelling in the CRs after the fit.

The VRs of the \twotau channel are presented in Table~\ref{tab:twotau:VRs}.
For the $Z(\tau\tau)$+jets background, the validation of the background extrapolation is performed from low-\HT to high-\HT, while keeping the upper bound on $\mTTT$ which is effective in selecting $Z(\tau\tau)$+jets events.
The validity of the top quark and $W(\tau\nu)$+jets background predictions obtained with alternative object selections are checked for selections with two reconstructed tau leptons. 
High values of $\mTTT$ are required to suppress $Z(\tau\tau)$+jets events as in the SRs, while upper bounds on \mttwo and \HT ensure there is no overlap between SRs and VRs.
The same set of CRs and VRs is used for the three SRs of the \twotau channel.

The resulting normalisation factors for both channels do not deviate from 1 by more than 25\%, 
except for \ttbar events with fake tau leptons in the \onetau channel, where the normalisation factor reaches 2 within large statistical uncertainty. 
The typical level of agreement between data and background distributions in the CRs after the fit can be seen in Figs.~\ref{fig:1tau:CRplots} and \ref{fig:2tau:CRplots}. Good modelling of kinematic distributions is observed in all CRs.
The comparison between the number of observed events and the predicted background yields in the VRs is displayed in Fig.~\ref{fig:VRpulls}. 
Agreement within approximately one standard deviation is observed.

\begin{table}[tb]
 \centering
 \caption{Overview of the control regions used in the \twotau channel.}
 {\renewcommand\arraystretch{1.2}
 \begin{tabular}{l|cc|cc|cc|c}
\hline
\hline
\twotau channel		&\multicolumn{2}{c|}{Kinematic CR}		&\multicolumn{2}{c|}{Fake tau CR} 		&\multicolumn{2}{c|}{True tau CR}	&	\\
		& $W$+jets 			&Top 			&$W$+jets 			&Top 			&$W$+jets 			&Top 		& $Z$+jets	\\
\hline
Trigger plateau	&\multicolumn{7}{c}{$\met> 180 \gev , \;\pt^\mathrm{jet_1} > 120 \gev$ }\\
\hline
Tau leptons		&\multicolumn{2}{c}{$N_\tau^\text{loose}=0$}	&\multicolumn{4}{|c|}{$N_\tau^\text{loose}=1$}	&$N_\tau^\text{loose}=2$, opposite charge\\
\hline
Light leptons	&\multicolumn{4}{c|}{$N_\mu=1$}	&\multicolumn{2}{c|}{$N_\mu=0$}&--	\\
\hline
\njet	&\multicolumn{2}{c|}{$\geq 3$}	&\multicolumn{2}{c|}{$\geq 2$} &\multicolumn{2}{c|}{$\geq 3$}& $\geq 2$ 	\\
\hline
$N_{b-\mathrm {jet}}$   & $=0$  & $ \geq 1$ &  $=0$  & $ \geq 1$ & $=0$  & $ \geq 1$ & $=0$ \\ 
\hline
Multi-jet rejection	&\multicolumn{6}{c|}{$0.4 < \Delta\phi(\text{jet}_{1,2},\ptmissv)< 2.9$} & $\Delta\phi(\text{jet}_{1,2},\ptmissv)>  0.4$\\
\hline
CR selection		&\multicolumn{6}{c|}{$\HT<\SI{1200}{\gev}$}							&$\HT<\SI{800}{\gev}$\\
\cline{2-8}
		&\multicolumn{2}{c|}{--}&\multicolumn{2}{c|}{$m_\mathrm{T}^\mu<100 \gev$}&\multicolumn{2}{c|}{$\mTtau<80 \gev$}&$\mTTT<150 \gev$\\
\hline\hline
 \end{tabular}
 }
\label{tab:twotau:CRs}
\end{table}

\begin{figure}[tb]
 \centering
  \subfigure[top quark fake-tau CR]{\includegraphics[width=.48\textwidth]%
                 {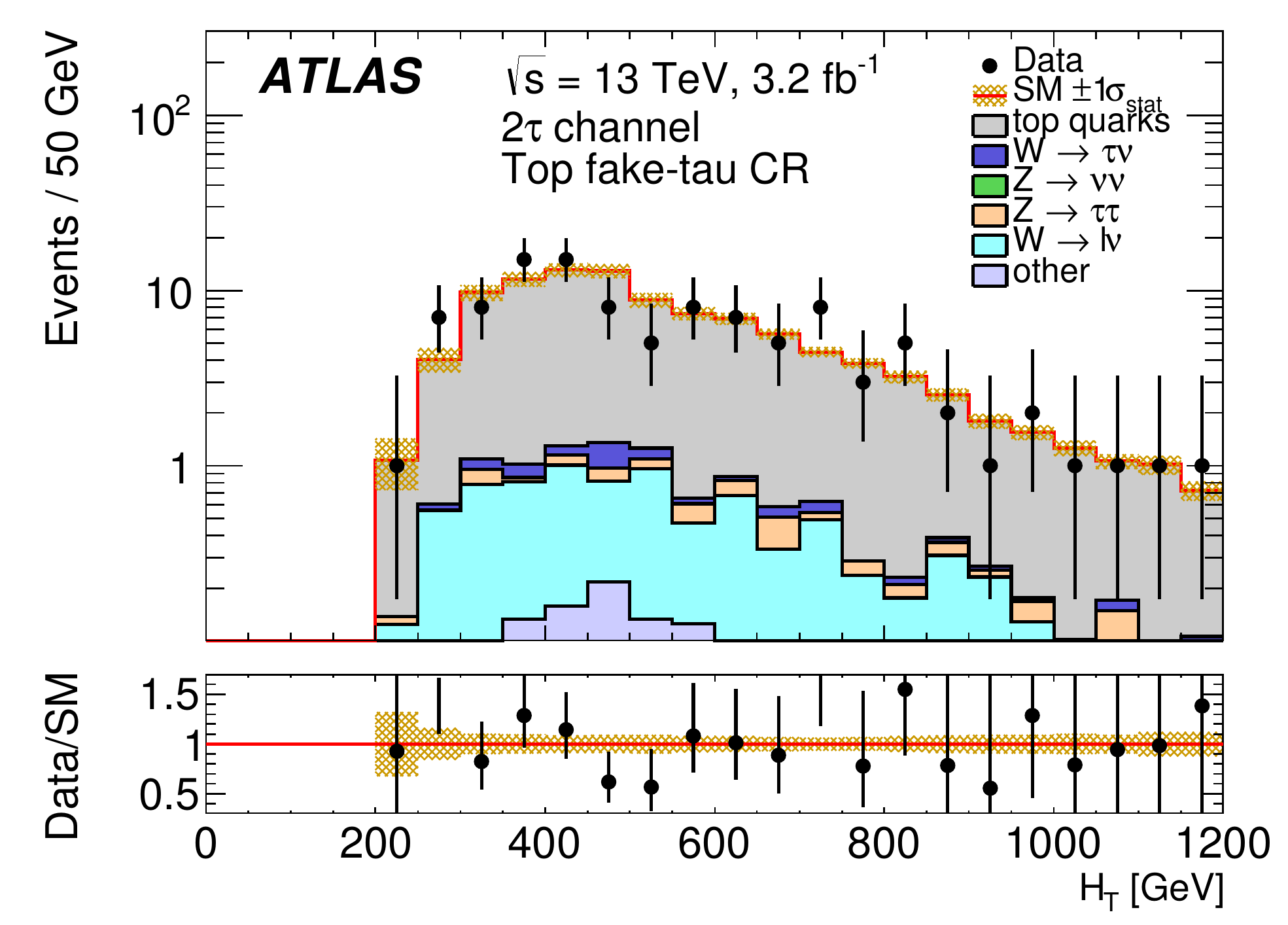} 
                 }
  \subfigure[$Z(\tau\tau)$+jets CR]{\includegraphics[width=.48\textwidth]%
                 {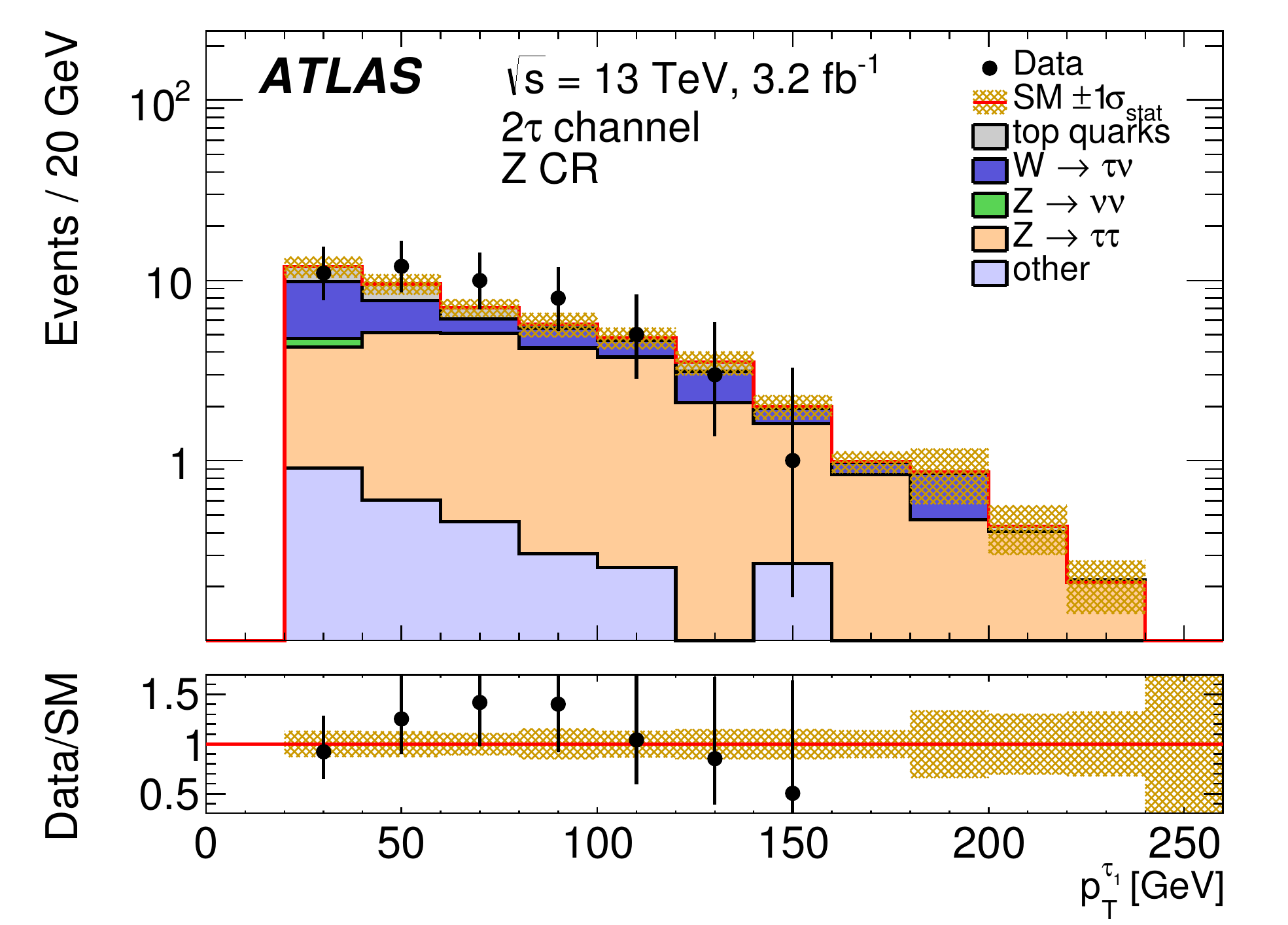}                  
                 }
  \caption{\HT distribution in the top quark fake-tau CR and transverse momentum of the leading tau lepton in the $Z(\tau\tau)$+jets CR of the \twotau channel after the fit, illustrating the overall background modelling in the control regions. 
  By construction, the total fitted background is equal to the number of observed events in each CR.
    The last bin includes overflow events. The shaded bands indicate the statistical uncertainties in the background predictions.
}
   \label{fig:2tau:CRplots}
\end{figure}

\begin{table}[htbp]
 \caption{Overview of the validation regions defined in the \twotau channel.}
 \centering

 {\renewcommand\arraystretch{1.2}
 \begin{tabular}{l|c|c|c}
\hline
\hline
\twotau channel		&VR $W$+jets 			&VR Top 	&VR $Z$+jets	\\
\hline
Trigger plateau	&\multicolumn{3}{c}{$\met> 180 \gev , \;\pt^\mathrm{jet_1} > 120 \gev$ }\\
\hline
Tau leptons		&\multicolumn{3}{c}{$N_\tau^\text{loose}\geq2$}	\\
\hline
\njet & \multicolumn{3}{c}{$\geq 2$} \\
\hline
$N_{b-\mathrm{jet}}$ & $ =0$	& $ \geq1$	& $=0$\\
\hline
Multi-jet rejection	&\multicolumn{3}{c}{$\Delta\phi(\text{jet}_{1,2},\ptmissv)\geq 0.4$}\\
\hline
VR selection		&\multicolumn{2}{c|}{$\HT<800 \gev $}	&$\HT> 800 \gev$\\
		&\multicolumn{2}{c|}{$\mTTT> 150 \gev $}&$\mTTT<150 \gev$\\
		&\multicolumn{2}{c|}{$\mttwo^{\tau\tau}<60 \gev$}&--\\
\hline\hline
 \end{tabular}
 }
\label{tab:twotau:VRs}
\end{table}

\begin{figure}[htb]
 \centering
  \includegraphics[width=1.0\textwidth]{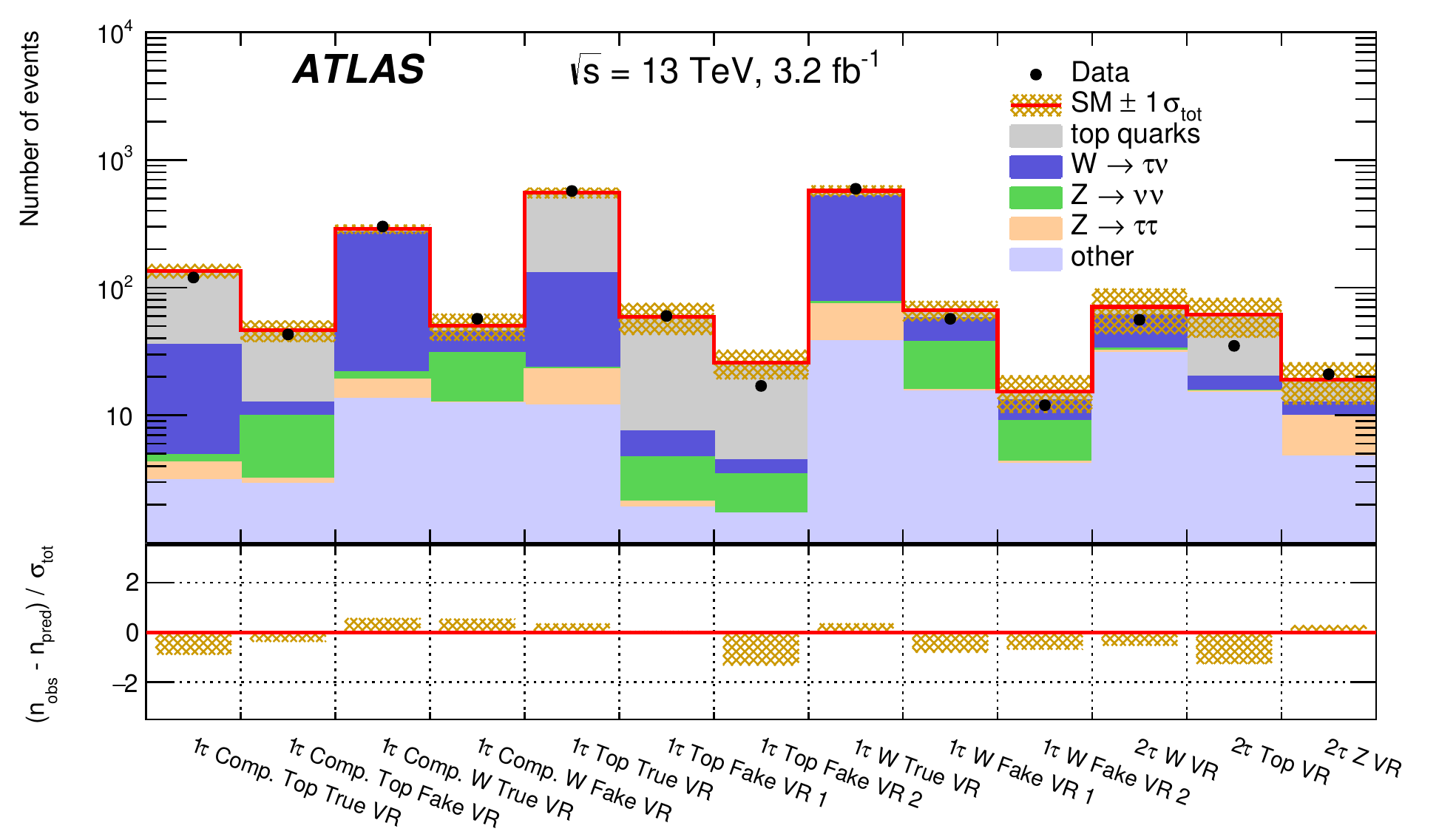}
 \caption{Number of observed events, $n_\mathrm{obs}$, and predicted background yields after the fit, $n_\mathrm{pred}$, in the validation regions of the \onetau and \twotau channels. 
The background predictions are scaled using normalisation factors derived in the control regions. 
 The shaded bands indicate the statistical uncertainties in the background predictions, and correspond to the $\sigma_\mathrm{tot}$ uncertainties used in the lower part of the figure.}
   \label{fig:VRpulls}
\end{figure}

\subsection{Multi-jet background}

The multi-jet background contributes to the selection when two conditions are simultaneously fulfilled: jets have to be mis-identified as tau leptons, and large missing transverse momentum must arise from jet energy mis-measurement.
This background is estimated from data, because final-state objects arising from mis-measurements are much more challenging to simulate than the reconstruction and identification of genuine objects. 
Moreover, the very large multi-jet production cross section at the LHC would imply simulating a prohibitively large number of multi-jet events.

The {\it jet smearing} method~\cite{SUSY-2011-20} employed in the \onetau channel proceeds in two steps.
First, multi-jet events with well-measured jets are selected in a data sample collected by single-jet triggers.
This is achieved by requiring $\met/(\sum E_\mathrm{T})^{1/3} < 5 \gev^{2/3}$, where the objects entering the $\sum E_\mathrm{T}$ term are those entering the \met calculation.
Selected events are required to have at least two jets, no light lepton and exactly one tau candidate satisfying the medium identification criteria.
The selection is dominated by multi-jet production, such that most tau candidates are jets mis-identified as tau leptons.
In a second step, jet energies are smeared according to the \pt-dependent jet energy resolution extracted from simulation.
The smearing is performed multiple times for each event, leading to a large pseudo-data set where \met originates from resolution effects and which includes an adequate fraction of jets mis-identified as tau leptons.

This method cannot be used in the \twotau channel because of the limited number of events with well-measured jets that contain at least two loose tau candidates.
Instead, a {\it fake rate} approach is adopted. 
The probability for jets to be mis-identified as tau leptons or muons, obtained from simulated dijet events, is applied to jets from an inclusive data sample collected by single-jet triggers and dominated by multi-jet events.

\section{Systematic uncertainties}
\label{sec:systematics}

For all simulated processes, theoretical and experimental systematic uncertainties are considered.
The former includes cross-section uncertainties, which are not relevant for the dominant backgrounds normalised to data, and generator modelling uncertainties.
The latter refers to all the uncertainties related to the reconstruction, identification, calibration and corrections applied to jets, tau leptons, electrons, muons and missing transverse momentum.
Specific uncertainties are evaluated for the multi-jet background, which is estimated from data.

Theoretical uncertainties are evaluated for all simulated samples. 
For backgrounds that are normalised in CRs, the uncertainty in the transfer factors, i.e. the ratio of the expected event yields in a SR or VR over the respective CR, is evaluated for all SRs and VRs. 
The difference between the nominal simulation and the systematically varied sample is used as an additional uncertainty. For backgrounds that are evaluated from simulation alone, i.e. the diboson background, a global normalisation uncertainty is added.
Uncertainties for $V$+jets samples generated with {\sc Sherpa} are estimated by up and down variations by factors of two and one-half in the renormalisation and factorisation scales, resummation scale (maximum scale of the additional emission to be resummed by the parton shower) and CKKW matching scale (matching between matrix elements and parton shower).
The effect of scale variations is parameterised at generator level as a function of the vector boson \pt and the number of particle jets.
For the top quark background, the nominal predictions from {\sc Powheg-Box}~+~{\sc Pythia6} are compared with predictions from alternative generators, and the differences are taken as systematic uncertainties.
The {\sc MG5\_aMC@NLO}~+~{\sc Herwig++} generators are used to evaluate uncertainties in the modelling of the hard scattering, parton shower and hadronisation.
The {\sc Powheg-Box}~+~{\sc Herwig++} generators are used to compute a specific uncertainty in the parton shower and hadronisation models.
An uncertainty in the ISR modelling is also assessed by varying the {\sc Powheg-Box} parameter which controls the transverse momentum of the first additional parton emission beyond the Born configuration.
In the case of the diboson background, a 6\% uncertainty in the cross section due to scale and PDF uncertainties is considered.
Uncertainties in signal cross sections are obtained by using different PDF sets and factorisation and renormalisation scales, as discussed in Section~\ref{sec:dataMC}.

Systematic uncertainties affecting jets arise from the jet energy scale and jet energy resolution~\cite{ATL-PHYS-PUB-2015-015}, as well as efficiency corrections for jet-vertex-tagging~\cite{ATLAS-CONF-2014-018} and $b$-tagging~\cite{ATLAS-CONF-2014-046}.
A set of \pt- and $\eta$-dependent uncertainties in the jet energy scale and resolution is estimated by varying the conditions used in the simulation. 
Another set of uncertainties accounts for the modelling of the residual pile-up dependence.
Additional uncertainties account for the jet flavour composition of samples that are used to derive in-situ energy scale corrections,
where jets are calibrated against well-measured objects. 
A punch-through uncertainty for jets not entirely contained in the calorimeters, as well as a single-hadron response uncertainty, are also included for high-\pt jets.
An overall uncertainty in the jet energy resolution is applied to jets in the simulation as a Gaussian energy smearing.

Systematic uncertainties affecting correctly identified tau leptons arise from the reconstruction, identification and tau-electron overlap-removal efficiencies, and the energy scale calibration~\cite{ATL-PHYS-PUB-2015-045}.
Most of the uncertainties are estimated by varying nominal parameters in the simulation: detector material, underlying event, hadronic shower model, pile-up and noise in the calorimeters.
The uncertainty in the energy scale also includes non-closure of the calibration found in simulation, a single-pion response uncertainty, 
and an uncertainty in the in-situ energy calibration of data with respect to simulation derived in $Z(\tau\tau)$ events with a hadronically decaying tau lepton and a muon in the final state.
In the case of signal samples, which undergo fast calorimeter simulation, a dedicated uncertainty takes into account the difference in performance between full and fast simulation.
The effect of mis-identified tau leptons is largely constrained by the background estimation approaches. Uncertainties arise due to the extrapolation from the CRs to the VRs and SRs. 
These are considered as part of the theory uncertainties, which account for the impact of hadronisation on the mis-identification of jets as tau leptons.

Systematic uncertainties affecting electrons and muons are related to the energy or momentum calibration, as well as efficiency corrections for the reconstruction, identification and isolation requirements.
These uncertainties have a negligible impact on the background predictions.

Systematic uncertainties in the missing transverse momentum originate from uncertainties in the energy or momentum calibration of jets, tau leptons, electrons, and muons, which are propagated to the \met calculation.
Additional uncertainties are related to the calculation of the track-based soft-term. 
These uncertainties are derived by studying the \pt balance in $Z(\mu\mu)$ events between the soft-term and the hard-term composed of all reconstructed objects.
Soft-term uncertainties include scale uncertainties along the hard-term axis, and resolution uncertainties along and perpendicular to the hard-term axis~\cite{ATL-PHYS-PUB-2015-023}.

A systematic uncertainty of the pile-up modelling is estimated by varying the distribution of the average number of interactions per bunch crossing in the simulation.
The range of the variation is determined by studying the correlation in data and simulation between the average number of interactions and the number of reconstructed primary vertices.
This uncertainty ranges from a few percent in the \onetau channel to about 15\% in the poorly populated SRs of the \twotau channel.

The uncertainty in the integrated luminosity is $\pm2.1$\%. 
It is derived, following a methodology similar to that detailed in Ref.~\cite{DAPR-2011-01}, from a calibration of the luminosity scale using $x$--$y$ beam-separation scans performed in August 2015.
This uncertainty only affects the diboson background prediction and the signal yields, as other backgrounds are normalised to the data.

The systematic uncertainty in the small multi-jet background contribution is estimated by considering alternative normalisation regions and different jet smearing parameters in the case of the \onetau channel.
An uncertainty of the order of 70\% is found for the \onetau channel, and a 100\% uncertainty is assigned in the \twotau channel.

The influence of the main systematic uncertainties in the total background predictions in the SRs of the \onetau and \twotau channels are summarised in Table~\ref{tab:systematics}.
The uncertainties reported in the table are derived assuming that no signal is present in the CRs.

The total uncertainties range between 13\% and 72\%. For all SRs, the largest uncertainties, between 7\% and 60\%, originate from the MC generator modelling of top quark events. Energy scale uncertainties affecting tau leptons and jets contribute significantly in all regions as well. 
Other uncertainties, e.g. in the $b$-tagging efficiency and the jet energy resolution, do not play a large role in most of the SRs.

\begin{table}[tb]
\begin{center}
\caption{Dominant systematic uncertainties in the total background predictions, in percent, for the signal regions of the \onetau (top) and \twotau (bottom) channels. 
The total systematic uncertainty accounts for other minor contributions not listed in this table, as well as correlations between the uncertainties.} 
\label{tab:systematics}
\setlength{\tabcolsep}{0.0pc}
\begin{tabular*}{\textwidth}{@{\extracolsep{\fill}}lrrr}
\noalign{\smallskip}\hline \hline\noalign{\smallskip}
Source of uncertainty	  & \onetau Compressed SR         & \onetau Medium-Mass SR            & \onetau High-Mass SR             \\
\noalign{\smallskip}\hline\noalign{\smallskip}
Top generator modelling & 	$ 8.1 $ &	$6.5 $ &	$12 $ \\
$V$+jets generator modelling & $1.5 $	&$6.4$ 	& $6.3$ \\
Jet energy scale & 	$2.0 $	&	$6.7$  & $0.4$ \\
Jet energy resolution & $0.6$	 & $0.2$	 & $0.7$	 \\
$b$-tagging efficiencies 	&  $1.9 $	& $3.2 $	& $7.7$ \\
Tau energy scale 	& $1.8 $ 	& $2.8$	&  $5.5$ \\ 
\hline
Total & $ 13 $ & $ 16 $ & $ 21$ \\
\hline \hline\noalign{\smallskip}
\\
\noalign{\smallskip}\hline \hline\noalign{\smallskip}
Source of uncertainty	  & \twotau Compressed SR         & \twotau High-Mass SR            & \twotau GMSB SR             \\
\noalign{\smallskip}\hline\noalign{\smallskip}
Top generator modelling 	& $60$	& $23$	& $22$	\\
$V$+jets generator modelling 	& $4.2$	& $6.3$ 	& $4.3$ 	\\
Jet energy scale 		& $14$ 	& $2.0$	& $6.0$ 	\\
Jet energy resolution 		& $8.1$	& $1.2$	& $4.3$	\\
$b$-tagging efficiencies 	& $8.8$	& $5.1$	& $7.7$ 	\\
Tau energy scale 		& $19$	& $13$	& $8.5$	\\ 
\hline
Total 				& $72 $	& $36$	& $35$	\\

\hline \hline\noalign{\smallskip}
\end{tabular*}

\end{center}
\end{table}

\FloatBarrier

\section{Results}
\label{sec:results}
Kinematic distributions for {\it extended SR selections} of the \onetau and \twotau channels are shown in Figs.~\ref{fig:1tau:ExtendedSRs} and ~\ref{fig:2tau:ExtendedSRs}, respectively.
These regions are defined by the same set of selection criteria as for the SRs, except that the criterion corresponding to the plotted variable is not applied. 
Data and background predictions, fitted to data in the control regions, are compared, and signal predictions are also shown for several benchmark models.  
Variables providing the most discrimination between signal and background are displayed:
$\mT^\tau$ distributions for the \onetau channel, and $\mT^\mathrm{sum}$, \mTTT and \HT distributions for the \twotau channel.
Reasonable agreement between data and background distributions is observed, given the low event yields remaining in data after these selections.
The numbers of observed events and expected background events in all SRs are reported in Tables~\ref{tab:onetau:yields_SR} and \ref{tab:twotau:yields_SR} for the \onetau and \twotau channels, respectively. 

\begin{figure}[tb]
 \centering
  \subfigure[]
{\includegraphics[width=.49\textwidth]%
                 {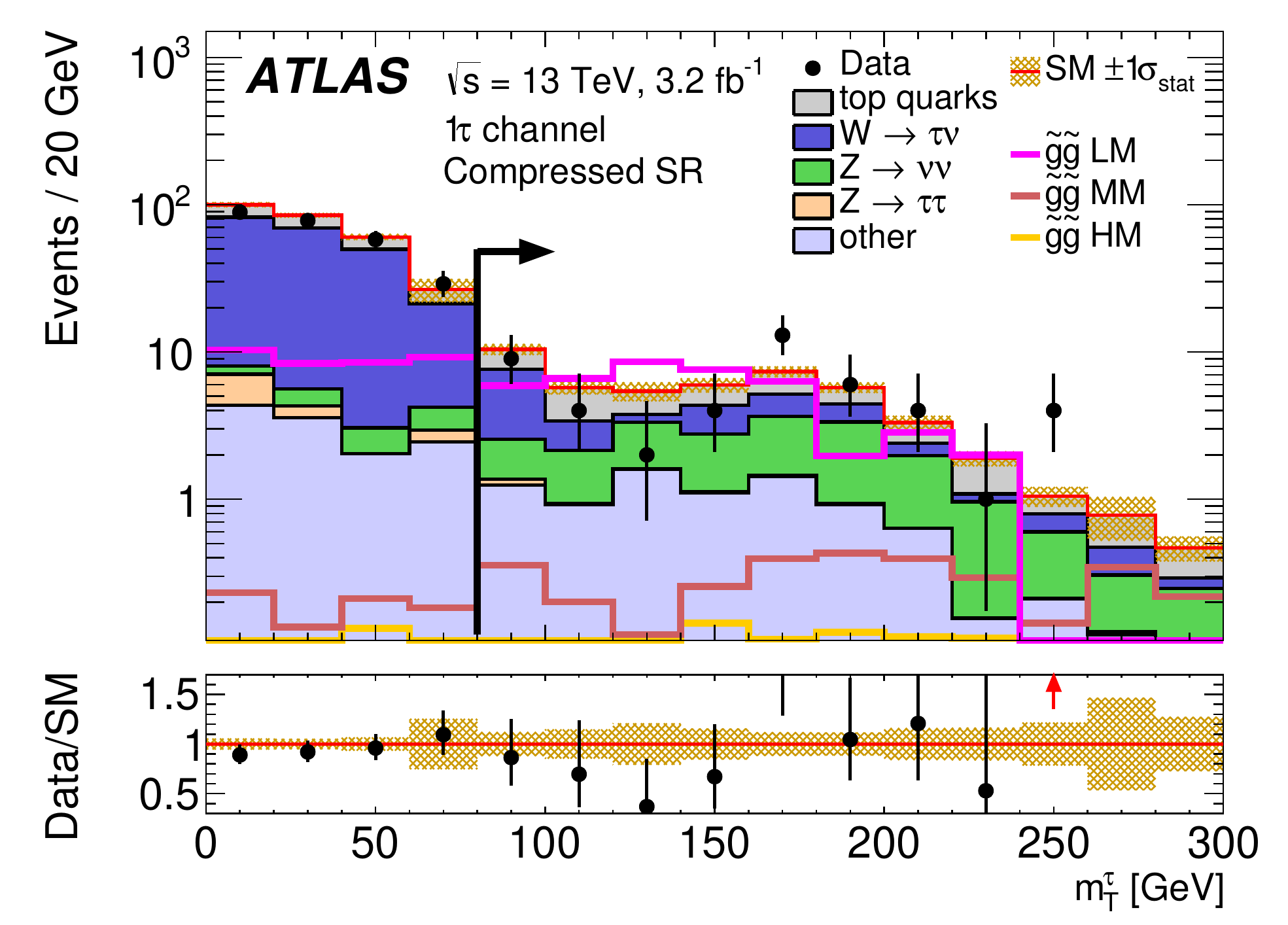}
                 }
  \subfigure[]
{\includegraphics[width=.49\textwidth]%
                 {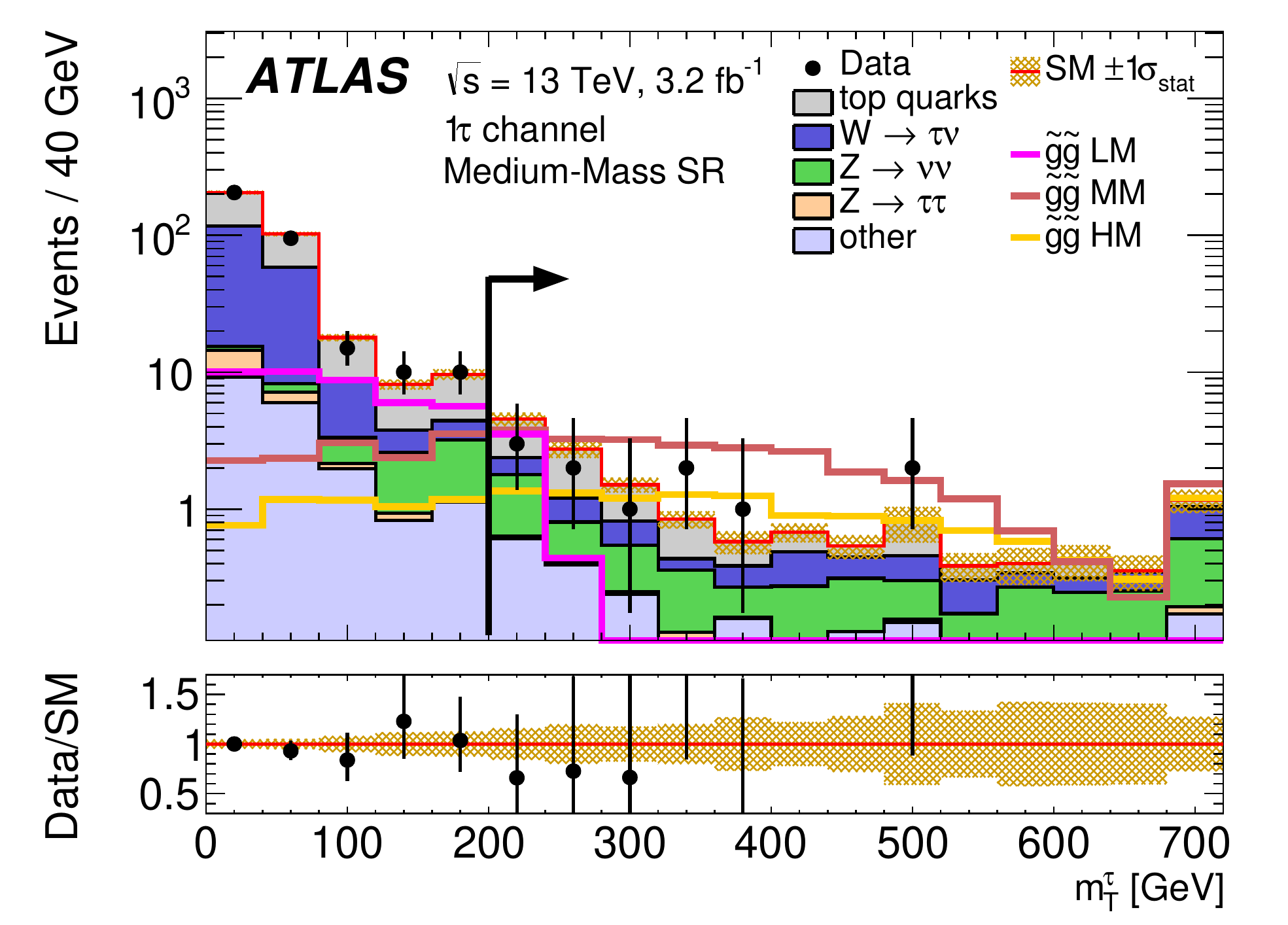}
                 }
   \subfigure[]
{\includegraphics[width=.49\textwidth]%
                 {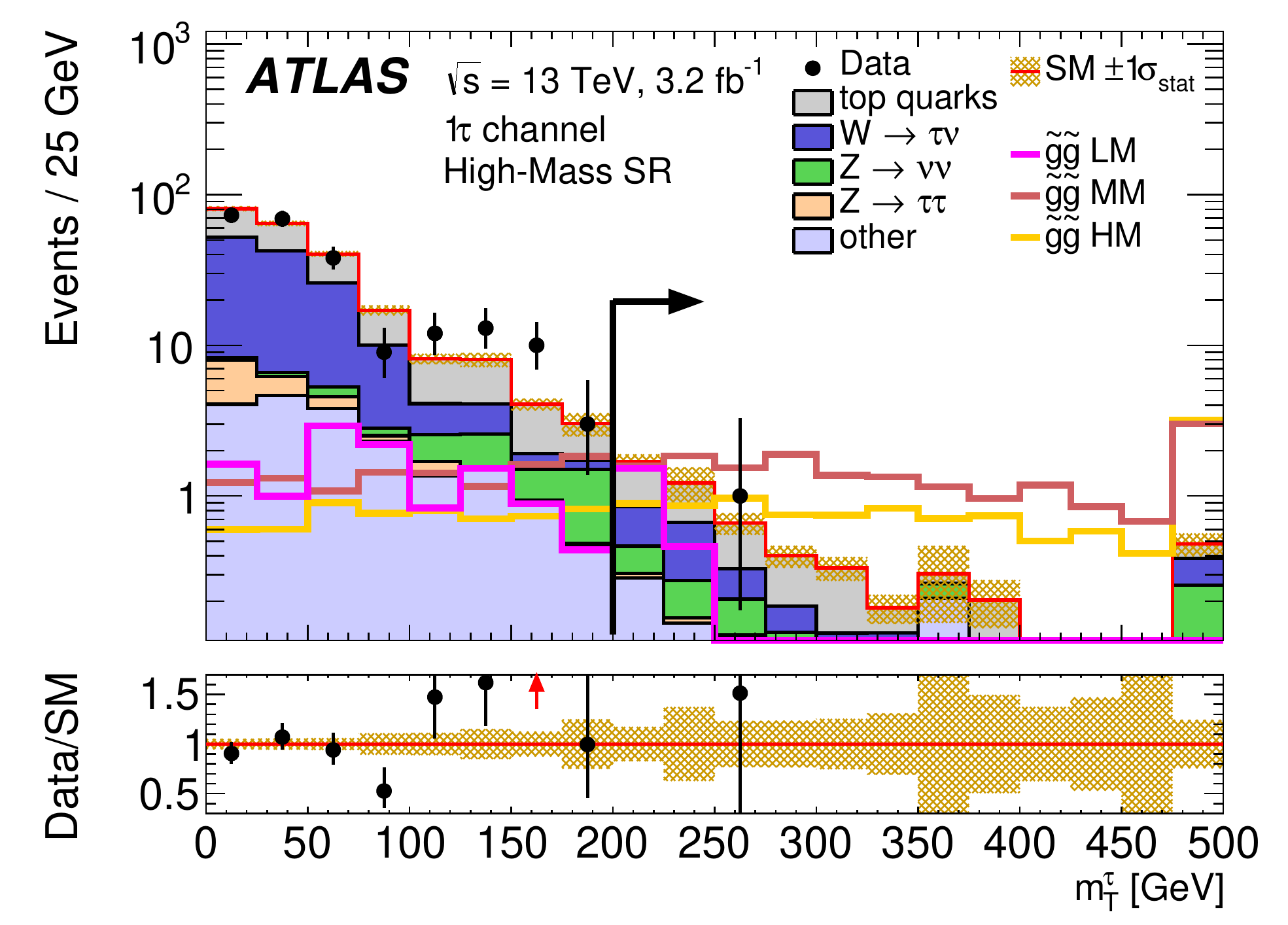}
                 }
  \caption{$m_\mathrm{T}^\tau$ distributions for ``extended SR selections'' of the \onetau channel, 
  for (a) the Compressed SR selection without the $\mT^\tau > 80\gev$ requirement,
  (b) the Medium-Mass SR selection without the $\mT^\tau > 200\gev$ requirement, and 
  (c) the High-Mass SR selection without the $\mT^\tau>200 \gev$ requirement.
  The last bin includes overflow events. The shaded bands indicate the statistical uncertainties in the background predictions.
    Red arrows in the Data/SM ratio indicate bins where the corresponding entry falls outside the plotted range.
	The signal region is indicated by the black arrow.
	Signal predictions are overlaid for several benchmark models, normalised to their predicted cross sections. 
	For the simplified model, ``LM'' refers to a low mass splitting, or compressed scenario, with $m_{\tilde g}=665 \gev$ and $m_{\tilde\chi_1^0}=585 \gev$; 
	``MM'' stands for a medium mass splitting, with $m_{\tilde g}=1145 \gev$ and $m_{\tilde\chi_1^0}=265 \gev$;
	``HM'' denotes a high mass splitting scenario, with $m_{\tilde g}=1305 \gev$ and $m_{\tilde\chi_1^0}=105 \gev$.
  }
   \label{fig:1tau:ExtendedSRs}
\end{figure}

\begin{figure}[tb]
 \centering
\subfigure[]
{\includegraphics[width=0.49\textwidth]
		{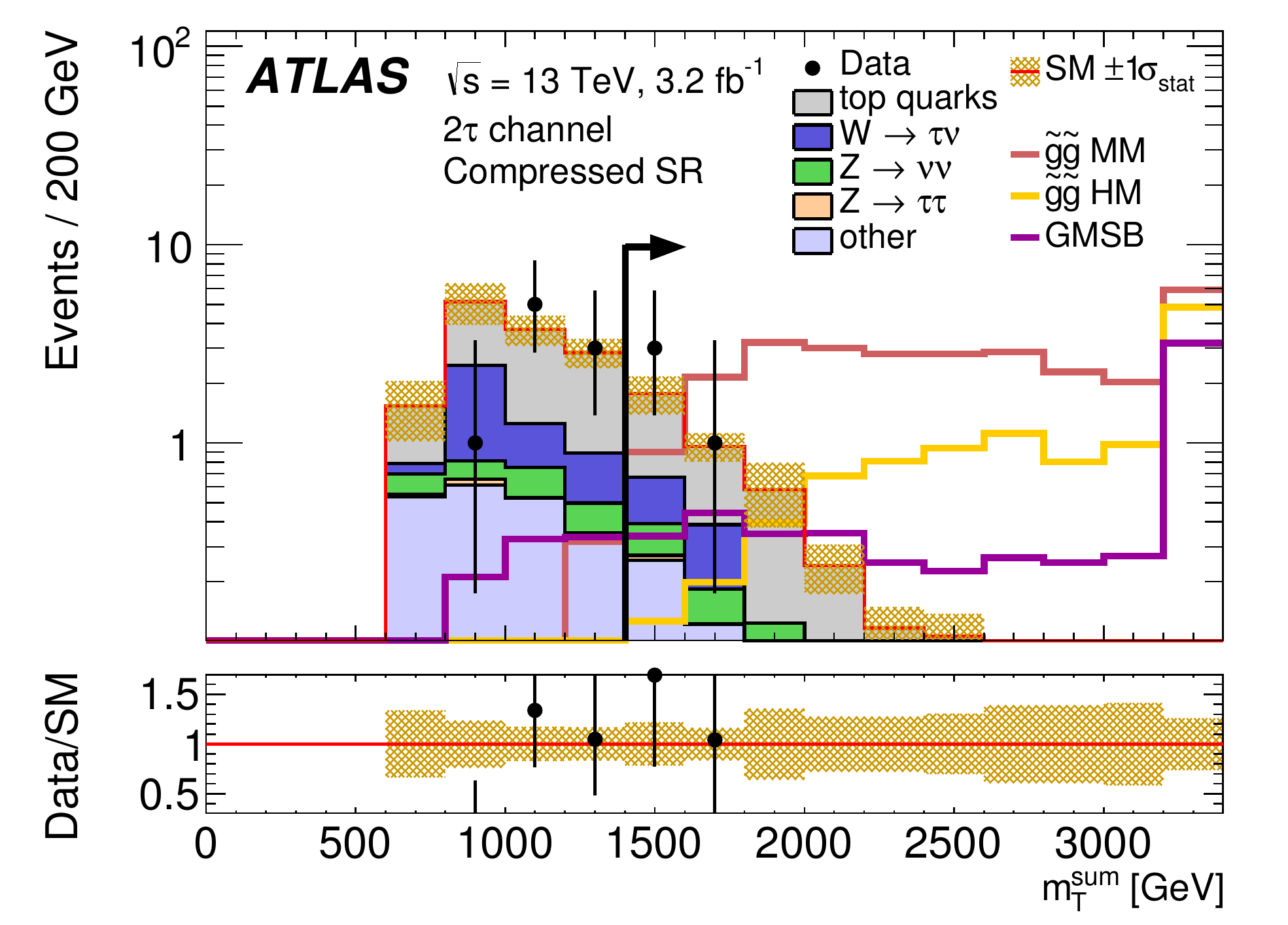}
		}
\subfigure[]
{\includegraphics[width=0.49\textwidth]{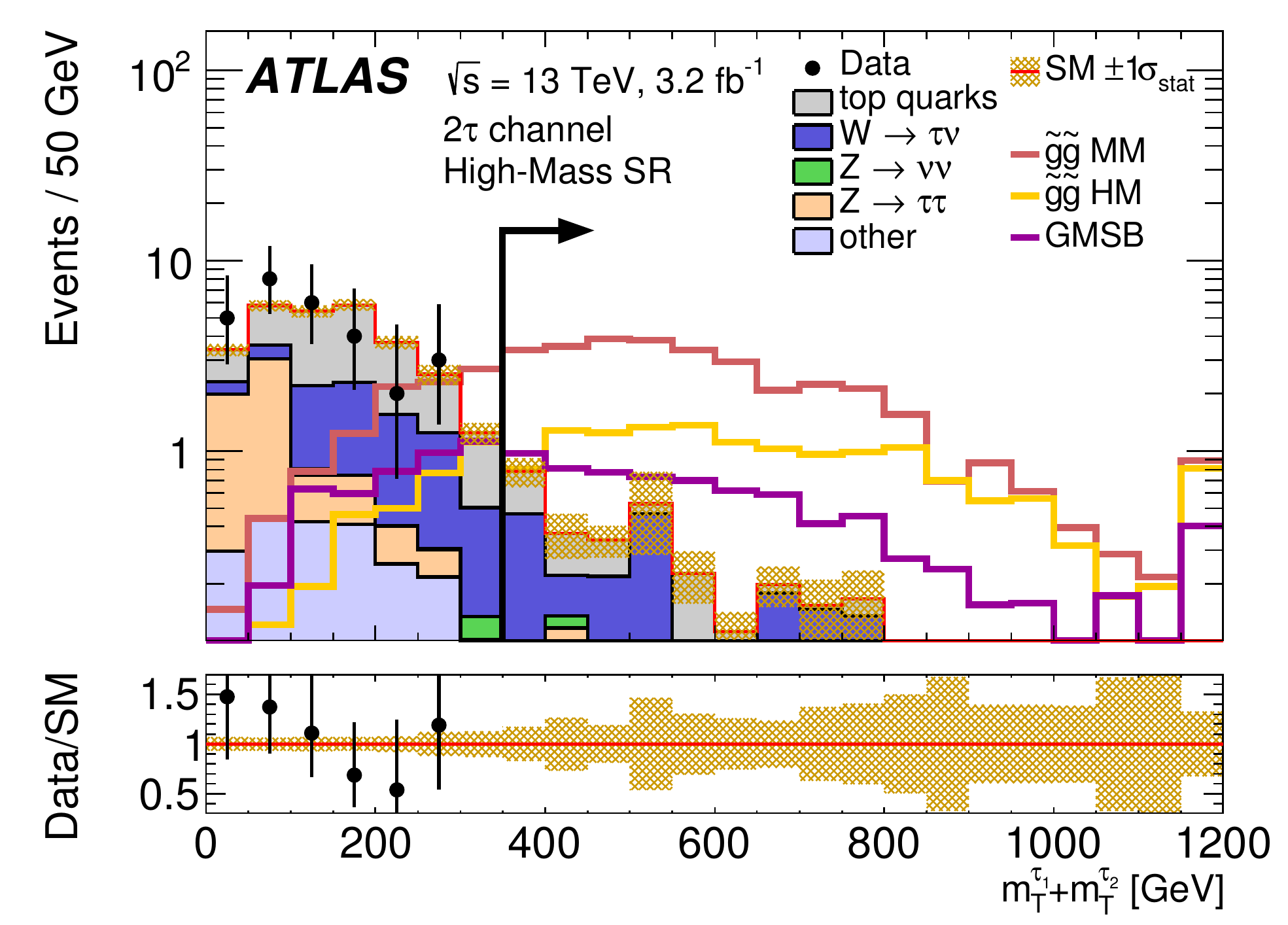}
		}
\subfigure[]
{\includegraphics[width=0.49\textwidth]		{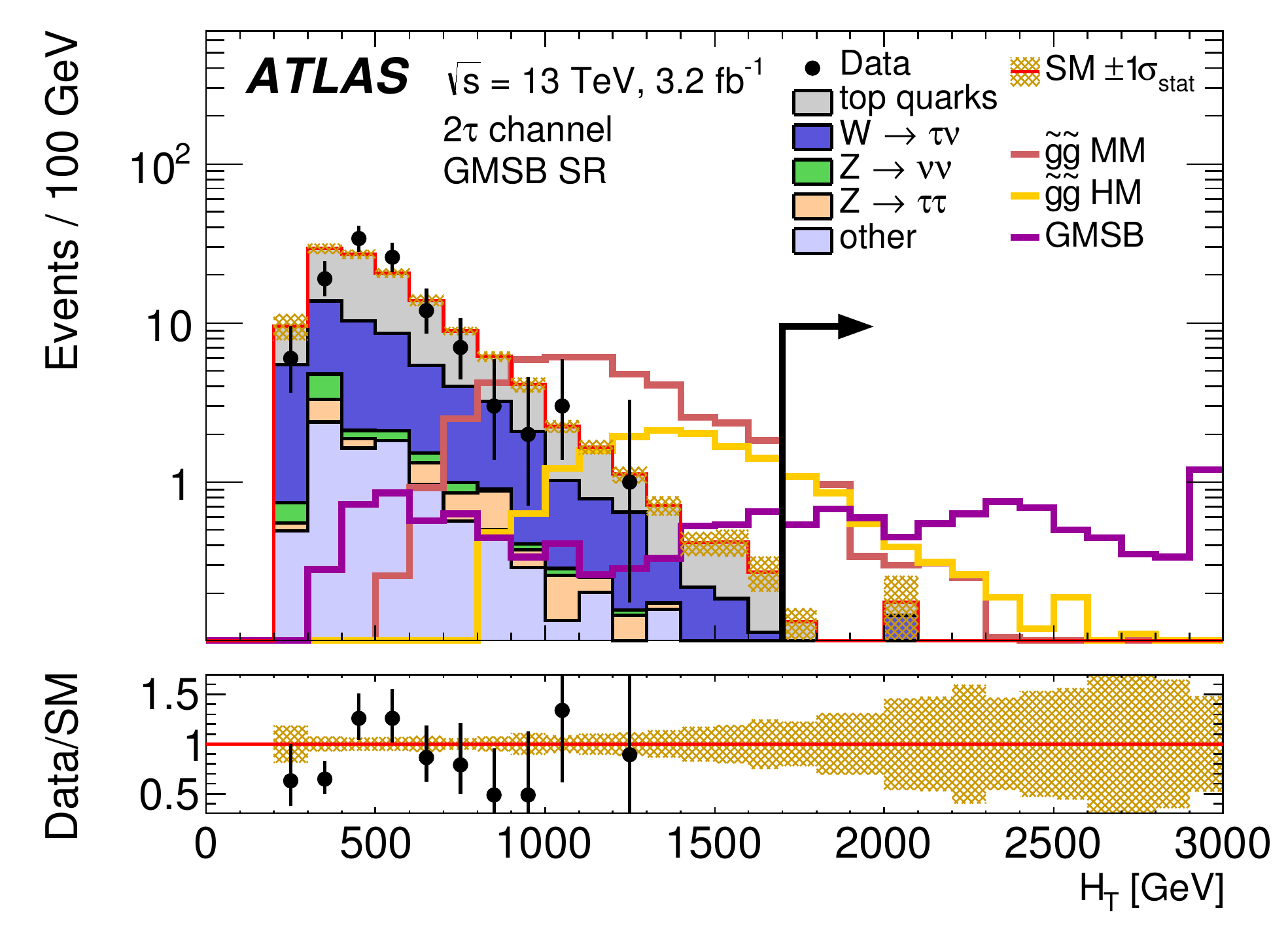}
		}
  \caption{Kinematic distributions for ``extended SR selections'' of the \twotau channel,
for (a) $\mT^\mathrm{sum}$ in the Compressed SR selection without the $\mT^\mathrm{sum}>1400 \gev$ requirement,
(b) \mTTT in the High-Mass SR selection without the $\mTTT>350\gev$ requirement, and 
(c) \HT in the GMSB SR selection without the $\HT > 1700 \gev$ requirement.
	The last bin includes overflow events. The shaded bands indicate the statistical uncertainties in the background predictions.
	The signal region is indicated by the black arrow.
	Signal predictions are overlaid for several benchmark models, normalised to their predicted cross sections. 
	For the simplified model, 
	``MM'' refers to a medium mass splitting, with $m_{\tilde g}=1145 \gev$ and $m_{\tilde\chi_1^0}=265 \gev$;
	``HM'' denotes a high mass splitting scenario, with $m_{\tilde g}=1305 \gev$ and $m_{\tilde\chi_1^0}=105 \gev$.
	The GMSB benchmark model corresponds to $\Lambda = 90\tev$ and $\tan\beta=40$. 
  }
   \label{fig:2tau:ExtendedSRs}
\end{figure}

No significant excess is observed in data over the SM predictions.
Therefore, upper limits are set at the 95\% confidence level (CL) on the number of hypothetical signal events, or equivalently, on the signal cross section.
The one-sided profile-likelihood-ratio test statistic is used to assess the compatibility of the observed data with the background-only and signal-plus-background hypotheses.
Systematic uncertainties are included in the likelihood function as nuisance parameters with Gaussian probability densities.
Following the standards used for LHC analyses, $p$-values are computed according to the $\mathrm{CL_s}$ prescription~\cite{0954-3899-28-10-313}.

Model-independent upper limits are calculated for each SR, assuming no signal contribution in the CRs.
The results are derived using profile-likelihood-ratio distributions obtained from pseudo-experiments.
Upper limits on signal yields are converted into limits on the visible cross section ($\sigma_\mathrm{vis}$) of BSM processes by dividing by the integrated luminosity of the data.
The visible cross section is defined as the product of production cross section, acceptance and selection efficiency. 
Results are summarised at the bottom of Tables~\ref{tab:onetau:yields_SR} and \ref{tab:twotau:yields_SR}.
The most stringent observed upper limits on the visible cross section are $1.17~\mathrm{fb}$ for the High-Mass SR of the \onetau channel and $1.07~\mathrm{fb}$ for the High-Mass and GMSB SRs of the \twotau channel.

\begin{table}[tb]
\caption{Number of observed events and predicted background yields in the three signal regions of the \onetau channel. 
The background prediction is scaled using normalisation factors derived in the control regions. All systematic and statistical uncertainties are included in the quoted uncertainties.
The bottom part of the table shows the observed and expected model-independent upper limits at the 95\% CL on the number of signal events, $S_{\rm obs}^{95}$  and $ S_{\rm exp}^{95}  $, respectively, 
 the corresponding observed upper limit on the visible cross section, $\langle{\rm \sigma_{vis}}\rangle_{\rm obs}^{95}$,
and the $\mathrm{CL_b}$ value, i.e. the confidence level observed for the background-only hypothesis.
}
\begin{center}
\setlength{\tabcolsep}{0.0pc}
{\small
\begin{tabular*}{\textwidth}{@{\extracolsep{\fill}}lrrr}
\noalign{\smallskip}\hline \hline\noalign{\smallskip}
\onetau channel	  & Compressed SR         & Medium-Mass SR            & High-Mass SR             \\[-0.05cm]
\noalign{\smallskip}\hline\noalign{\smallskip}
Data          & $47$ & $11$              & $1$                    \\
\noalign{\smallskip}\hline\noalign{\smallskip}
Total background        & $49.2 \pm 6.2$   & $15.0 \pm 2.4$          & $5.7 \pm 1.2$              \\
\noalign{\smallskip}\hline\noalign{\smallskip}
	Top  &   $14.3 \pm 4.5$   & $6.0 \pm 1.3$          & $2.49 \pm 0.87$ 	\\
	$W(\tau\nu)$+jets  & $12.1 \pm 1.3$  & $2.78 \pm 0.62$          & $1.17 \pm 0.33$ \\
         $Z(\nu\nu)$+jets        & $13.9 \pm 2.3$    &$3.8 \pm 1.1$          & $0.83 \pm 0.21$            \\
	$V$+jets, other & $6.24 \pm 0.90$ 	& $1.44 \pm 0.32$          & $0.75 \pm 0.23$  \\
         Diboson         & $1.85 \pm 0.23$  & $0.76 \pm 0.16$          & $0.20 \pm 0.03$             \\
         Multi-jet         & $0.74 \pm 0.54$ & $0.19 \pm 0.18$          & $0.24 \pm 0.17$          \\

\noalign{\smallskip}\hline \hline\noalign{\smallskip}
$S_{\rm obs}^{95}$  ($ S_{\rm exp}^{95}  $) &$16.7$	 (${18.4}^{+6.9}_{-5.0}$) &   $7.5$ (${9.7}^{+3.5}_{-2.5}$)  &$3.8$ (${6.1}^{+2.1}_{-1.5}$) \\
$\langle{\rm \sigma_{vis}}\rangle_{\rm obs}^{95}$ [fb] & 	$5.19$	& $2.34$	& $1.17$\\
$\mathrm{CL_b}$&0.41&0.23&0.02\\
\hline\hline
\end{tabular*}
}
\end{center}
\label{tab:onetau:yields_SR}
\end{table}

\begin{table}[tb]
\caption{Number of observed events and predicted background yields in the three signal regions of the \twotau channel. 
The background prediction is scaled using normalisation factors derived in the control regions. All systematic and statistical uncertainties are included in the quoted uncertainties.
The bottom part of the table shows the observed and expected model-independent upper limits at the 95\% CL on the number of signal events, $S_{\rm obs}^{95}$  and $ S_{\rm exp}^{95}  $, respectively, 
 the corresponding observed upper limit on the visible cross section, $\langle{\rm \sigma_{vis}}\rangle_{\rm obs}^{95}$,
and the $\mathrm{CL_b}$ value, i.e. the confidence level observed for the background-only hypothesis.
}
\begin{center}
\setlength{\tabcolsep}{0.0pc}
{\small
\begin{tabular*}{\textwidth}{@{\extracolsep{\fill}}lrrr}
\noalign{\smallskip}\hline\hline\noalign{\smallskip}
\twotau channel           &  Compressed SR         & High-Mass SR            &  GMSB SR             \\[-0.05cm]
\noalign{\smallskip}\hline\noalign{\smallskip}
Data        & $4$              & $0$              & $0$                       \\
\noalign{\smallskip}\hline\noalign{\smallskip}
Total background        & $4.2 \pm 3.0$          & $3.2 \pm 1.2$          & $0.69 \pm 0.24$              \\
\noalign{\smallskip}\hline\noalign{\smallskip}
       Top       & $2.5_{-2.5}^{+2.9}$          & $0.87 \pm 0.78$          & $0.20 \pm 0.20$             \\
    $W(\tau\nu)$+jets          & $0.51 \pm 0.38$          & $1.75 \pm 0.65 $          & $0.31 \pm 0.14$             \\
    $Z(\tau\tau)$+jets       & $0.04 \pm 0.02$          & $0.13 \pm 0.06$          & $0.04 \pm 0.02$             \\
    $Z(\nu\nu)$+jets         & $0.28 \pm 0.12$          & $0.07 \pm 0.03$          & $0.02 \pm 0.01$           \\
     $W(\ell\nu)$+jets            & $0.37 \pm 0.34$          & $0.12 \pm 0.07$          & $0.02 \pm 0.01$           \\
    Diboson      & $0.25 \pm 0.10$          & $0.21 \pm 0.08$          & $0.06 \pm 0.02$        \\
   Multi-jet          & $0.21 \pm 0.21$          & $0.07 \pm 0.07$          & $0.06 \pm 0.06$             \\
\noalign{\smallskip}\hline \hline\noalign{\smallskip}

$S_{\rm obs}^{95}$  ($ S_{\rm exp}^{95}  $) &$ 8.2$	 (${8.0}^{+2.1}_{-2.0}$) &	$3.4$ (${4.8}^{+1.4}_{-1.0}$)  &$3.4$ (${3.7}^{+0.5}_{-0.2}$) \\
$\langle{\rm \sigma_{vis}}\rangle_{\rm obs}^{95}$ [fb] & 	$2.55$	& $1.07$	& $1.07$\\
$\mathrm{CL_b}$&0.53&0.12&0.53\\

\hline\hline
\end{tabular*}
}
\end{center}
\label{tab:twotau:yields_SR}
\end{table}

\clearpage

The results are interpreted in the context of the simplified model of gluino pair production and the GMSB model.
In the case of model-dependent interpretations, the signal contribution in the control regions is included in the calculation of upper limits, and asymptotic properties of test-statistic distributions are used~\cite{Cowan:2010js}.
Exclusion contours at the 95\% CL are derived in the $(m_{\tilde g},m_{\tilde\chi_1^0})$ plane for the simplified model and in the $(\Lambda,\tan\beta)$ plane for the GMSB model.
Results are shown in Figs.~\ref{fig:LimitsSimplifiedModel} and \ref{fig:LimitsGMSB}.
The solid lines and the dashed lines correspond to the observed and median expected limits, respectively.
The band shows the one-standard-deviation spread of the expected limits around the median, which originates from statistical and systematic uncertainties in the background and signal.
The theoretical uncertainty in the signal cross section is not included in the band. Its effect on the observed limits is shown separately as the dotted lines.
For the simplified model, exclusion contours are shown for the \onetau and \twotau channels and their combination.
The combination is performed by selecting, for each signal scenario, the SR with the lowest expected $\mathrm{CL_s}$ value.
The combination retains the Compressed SR of the \onetau channel in the region where the LSP mass is close to the gluino mass, and favours the High-Mass SR of the \twotau channel when the mass splitting is large.
For the GMSB model, limits are shown for the GMSB SR of the \twotau channel.
The stronger limits at high values of $\tan\beta$ are explained by the nature of the NSLP, which is the lightest scalar tau in this region.
For both models, the exclusion limits obtained with $3.2~\ifb$ of collision data at $\sqrt{s}=13 \tev$ significantly improve upon the previous ATLAS results~\cite{SUSY-2013-10,SUSY-2014-06} established with $20.3~\ifb$ of $8 \tev$ data.

\begin{figure}[tb]
 \centering
  {\includegraphics[width=0.88\textwidth]%
                 {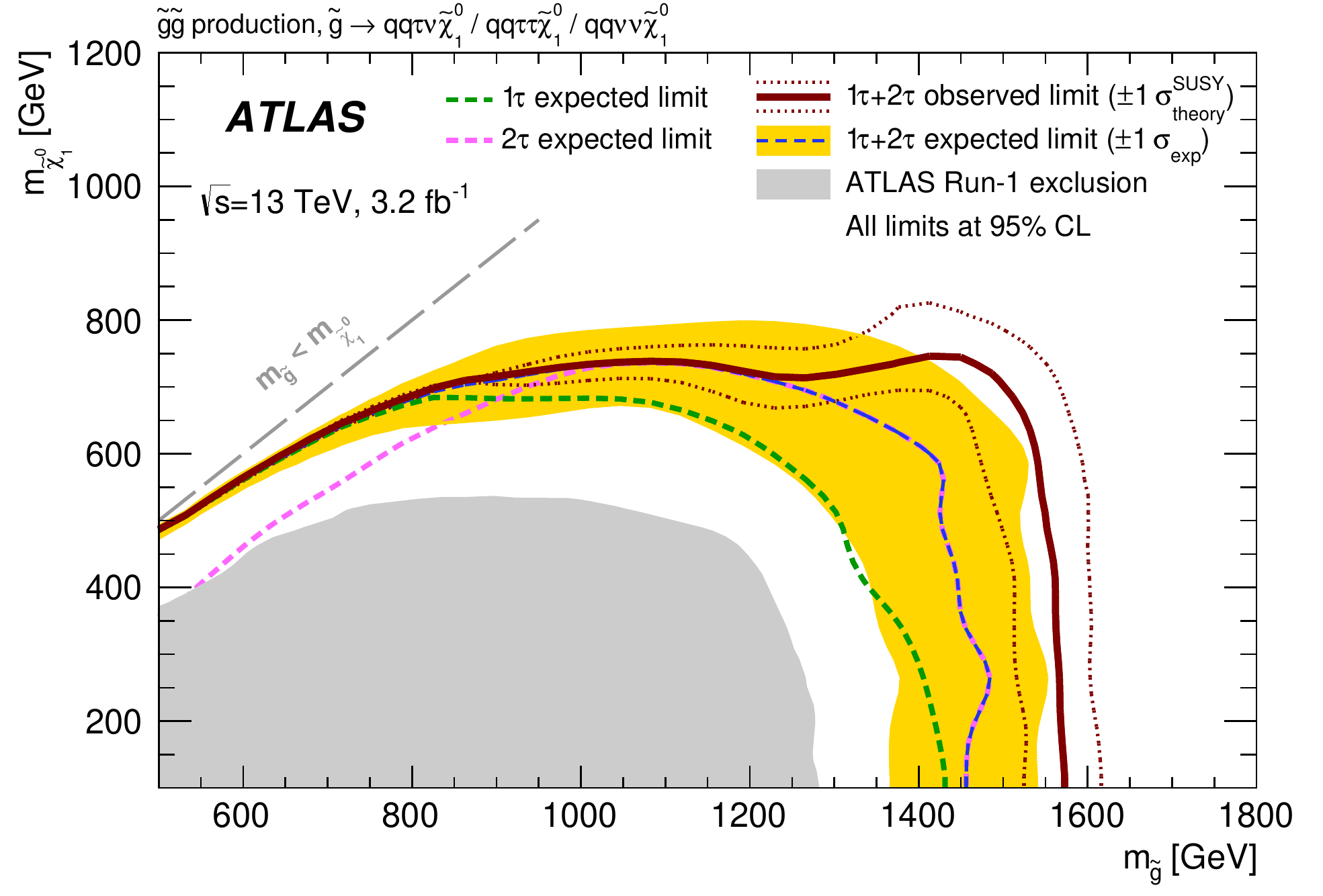}
		}
  \caption{Exclusion contours at the 95\% confidence level for the simplified model of gluino pair production, based on results from the \onetau and \twotau channels.
The red solid line and the blue dashed line correspond to the observed and median expected limits, respectively, for the combination of the two channels.
The yellow band shows the one-standard-deviation spread of expected limits around the median.
The effect of the signal cross-section uncertainty in the observed limits is shown as red dotted lines.
Additionally, expected limits are shown for the \onetau and \twotau channels individually as dashed green and magenta lines, respectively.
The previous ATLAS result~\cite{SUSY-2014-06} obtained with $20.3~\ifb$ of $8 \tev$ data is shown as the grey filled area.}
   \label{fig:LimitsSimplifiedModel}
\end{figure}

\begin{figure}[tb]
 \centering
  {\includegraphics[width=0.88\textwidth]%
                 {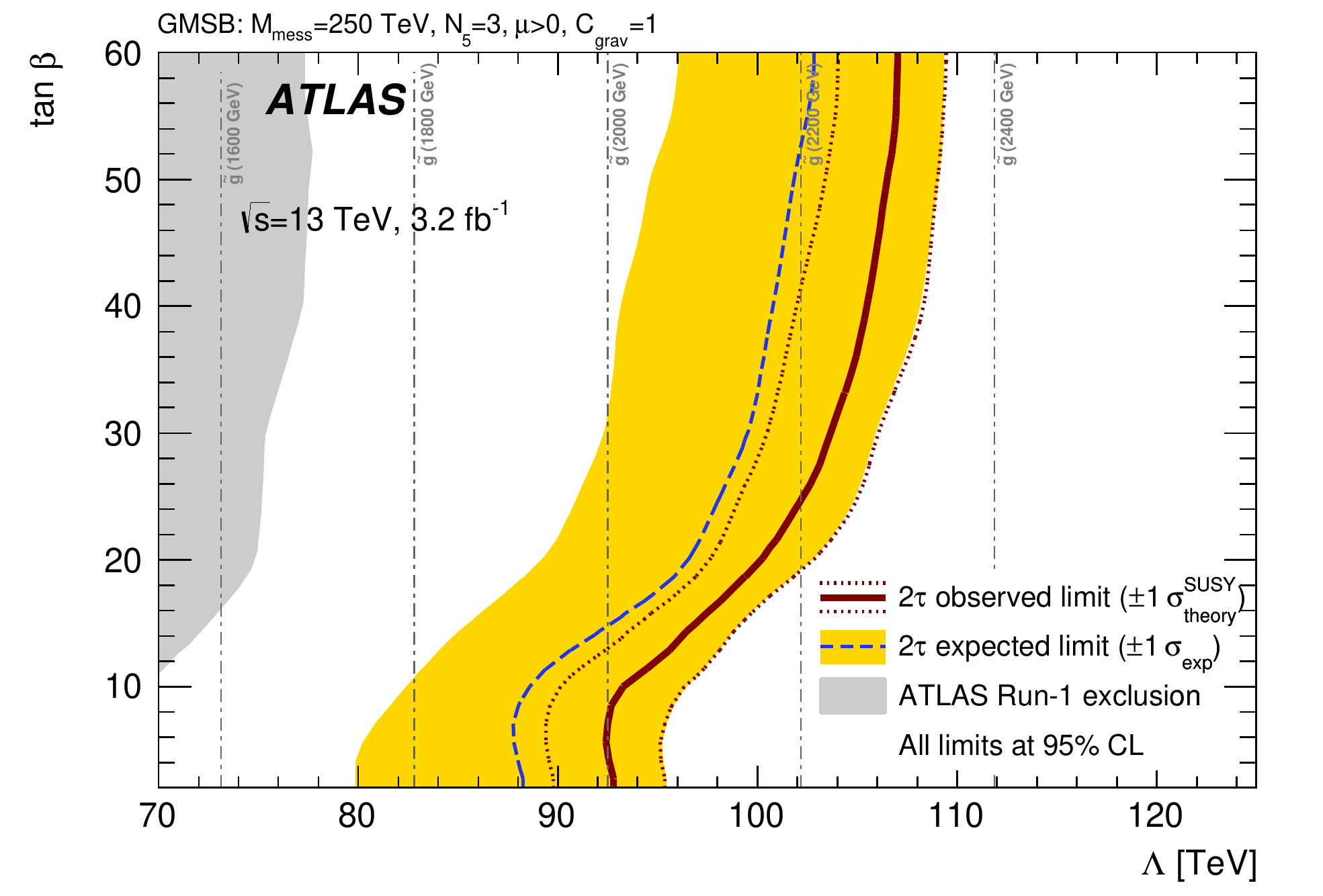}
		}
  \caption{Exclusion contours at the 95\% confidence level for the gauge-mediated supersymmetry-breaking model, based on results from the \twotau channel.
The red solid line and the blue dashed line correspond to the observed and median expected limits, respectively.
The yellow band shows the one-standard-deviation spread of expected limits around the median.
The effect of the signal cross-section uncertainty in the observed limits is shown as red dotted lines.
The previous ATLAS result~\cite{SUSY-2013-10} obtained with $20.3~\ifb$ of $8 \tev$ data is shown as the grey filled area.}
   \label{fig:LimitsGMSB}
\end{figure}

\section{Summary}
\label{sec:summary}

A search for squarks and gluinos has been performed in events with hadronically decaying tau leptons, jets and missing transverse momentum, using 3.2~\ifb\ of $pp$ collision data at $\sqrt{s}=13$~\tev\ recorded by the ATLAS detector at the LHC in 2015.
Two channels, with either one tau lepton or at least two tau leptons, are separately optimised.
The numbers of observed events in the different signal regions are in agreement with the Standard Model predictions.
Results are interpreted in the context of a gauge-mediated supersymmetry breaking model and a simplified model of gluino pair production with tau-rich cascade decay.
In the GMSB model, limits are set on the SUSY-breaking scale $\Lambda$ as a function of $\tan\beta$. 
Values of $\Lambda$ below $92 \tev$ are excluded at the 95\% CL, corresponding to gluino masses below $2000 \gev$.
A stronger exclusion is achieved for large values of $\tan\beta$, where $\Lambda$ and gluino mass values are excluded up to $107 \tev$ and $2300 \gev$, respectively.
In the simplified model, gluino masses are excluded up to $1570 \gev$ for neutralino masses around $100 \gev$,  neutralino masses up to $700 \gev$ are excluded for all gluino masses between $800 \gev$ and $1500 \gev$, while the strongest neutralino-mass exclusion of $750 \gev$ is achieved for gluino masses around $1400 \gev$.
A dedicated signal region provides good sensitivity to scenarios with a small mass difference between the gluino and the neutralino LSP.

\clearpage
\section*{Acknowledgements}

We thank CERN for the very successful operation of the LHC, as well as the
support staff from our institutions without whom ATLAS could not be
operated efficiently.

We acknowledge the support of ANPCyT, Argentina; YerPhI, Armenia; ARC, Australia; BMWFW and FWF, Austria; ANAS, Azerbaijan; SSTC, Belarus; CNPq and FAPESP, Brazil; NSERC, NRC and CFI, Canada; CERN; CONICYT, Chile; CAS, MOST and NSFC, China; COLCIENCIAS, Colombia; MSMT CR, MPO CR and VSC CR, Czech Republic; DNRF and DNSRC, Denmark; IN2P3-CNRS, CEA-DSM/IRFU, France; GNSF, Georgia; BMBF, HGF, and MPG, Germany; GSRT, Greece; RGC, Hong Kong SAR, China; ISF, I-CORE and Benoziyo Center, Israel; INFN, Italy; MEXT and JSPS, Japan; CNRST, Morocco; FOM and NWO, Netherlands; RCN, Norway; MNiSW and NCN, Poland; FCT, Portugal; MNE/IFA, Romania; MES of Russia and NRC KI, Russian Federation; JINR; MESTD, Serbia; MSSR, Slovakia; ARRS and MIZ\v{S}, Slovenia; DST/NRF, South Africa; MINECO, Spain; SRC and Wallenberg Foundation, Sweden; SERI, SNSF and Cantons of Bern and Geneva, Switzerland; MOST, Taiwan; TAEK, Turkey; STFC, United Kingdom; DOE and NSF, United States of America. In addition, individual groups and members have received support from BCKDF, the Canada Council, CANARIE, CRC, Compute Canada, FQRNT, and the Ontario Innovation Trust, Canada; EPLANET, ERC, FP7, Horizon 2020 and Marie Sk{\l}odowska-Curie Actions, European Union; Investissements d'Avenir Labex and Idex, ANR, R{\'e}gion Auvergne and Fondation Partager le Savoir, France; DFG and AvH Foundation, Germany; Herakleitos, Thales and Aristeia programmes co-financed by EU-ESF and the Greek NSRF; BSF, GIF and Minerva, Israel; BRF, Norway; Generalitat de Catalunya, Generalitat Valenciana, Spain; the Royal Society and Leverhulme Trust, United Kingdom.

The crucial computing support from all WLCG partners is acknowledged
gratefully, in particular from CERN and the ATLAS Tier-1 facilities at
TRIUMF (Canada), NDGF (Denmark, Norway, Sweden), CC-IN2P3 (France),
KIT/GridKA (Germany), INFN-CNAF (Italy), NL-T1 (Netherlands), PIC (Spain),
ASGC (Taiwan), RAL (UK) and BNL (USA) and in the Tier-2 facilities
worldwide.

\clearpage

\bibliographystyle{bibtex/bst/atlasBibStyleWithTitle} 
\bibliography{own,ATLAS,ConfNotes,PubNotes}
\clearpage

\begin{flushleft}
{\Large The ATLAS Collaboration}

\bigskip

M.~Aaboud$^\textrm{\scriptsize 136d}$,
G.~Aad$^\textrm{\scriptsize 87}$,
B.~Abbott$^\textrm{\scriptsize 114}$,
J.~Abdallah$^\textrm{\scriptsize 65}$,
O.~Abdinov$^\textrm{\scriptsize 12}$,
B.~Abeloos$^\textrm{\scriptsize 118}$,
R.~Aben$^\textrm{\scriptsize 108}$,
O.S.~AbouZeid$^\textrm{\scriptsize 138}$,
N.L.~Abraham$^\textrm{\scriptsize 152}$,
H.~Abramowicz$^\textrm{\scriptsize 156}$
% ...
\bigskip
\\
% ...
$^{12}$ Institute of Physics, Azerbaijan Academy of Sciences, Baku, Azerbaijan\\
% ...
$^{65}$ University of Iowa, Iowa City IA, United States of America\\
% ...
$^{87}$ CPPM, Aix-Marseille Universit{\'e} and CNRS/IN2P3, Marseille, France\\
% ...
$^{108}$ Nikhef National Institute for Subatomic Physics and University of Amsterdam, Amsterdam, Netherlands\\
% ...
$^{114}$ Homer L. Dodge Department of Physics and Astronomy, University of Oklahoma, Norman OK, United States of America\\
% ...
$^{118}$ LAL, Univ. Paris-Sud, CNRS/IN2P3, Universit{\'e} Paris-Saclay, Orsay, France\\
% ...
$^{136}$ $^{(a)}$ Facult{\'e} des Sciences Ain Chock, R{\'e}seau Universitaire de Physique des Hautes Energies - Universit{\'e} Hassan II, Casablanca; $^{(b)}$ Centre National de l'Energie des Sciences Techniques Nucleaires, Rabat; $^{(c)}$ Facult{\'e} des Sciences Semlalia, Universit{\'e} Cadi Ayyad, LPHEA-Marrakech; $^{(d)}$ Facult{\'e} des Sciences, Universit{\'e} Mohamed Premier and LPTPM, Oujda; $^{(e)}$ Facult{\'e} des sciences, Universit{\'e} Mohammed V, Rabat, Morocco\\
% ...
$^{138}$ Santa Cruz Institute for Particle Physics, University of California Santa Cruz, Santa Cruz CA, United States of America\\
% ...
$^{152}$ Department of Physics and Astronomy, University of Sussex, Brighton, United Kingdom\\
% ...
$^{156}$ Raymond and Beverly Sackler School of Physics and Astronomy, Tel Aviv University, Tel Aviv, Israel\\
% ...
$^{a}$ Also at Department of Physics, King's College London, London, United Kingdom\\
$^{b}$ Also at Institute of Physics, Azerbaijan Academy of Sciences, Baku, Azerbaijan\\
$^{c}$ Also at Novosibirsk State University, Novosibirsk, Russia\\
$^{d}$ Also at TRIUMF, Vancouver BC, Canada\\
$^{e}$ Also at Department of Physics {\&} Astronomy, University of Louisville, Louisville, KY, United States of America\\
% ...
\end{flushleft}

\clearpage
\appendix

\end{document}